\documentclass[preprint,aps,prd,superscriptaddress,nofootinbib]{revtex4-1}

\pdfoutput = 1

\usepackage[english]{babel}
\usepackage[utf8x]{inputenc}
\usepackage[T1]{fontenc}
\usepackage{ae} 
\usepackage{aecompl}
\usepackage{natbib}

\usepackage{epsfig,amsmath,amssymb,verbatim,mathrsfs,hyperref,graphicx}

\newcommand{\beq}{\begin{eqnarray}}
\newcommand{\eeq}{\end{eqnarray}}
\newcommand{\bea}{\begin{eqnarray}}
\newcommand{\eea}{\end{eqnarray}}

\newcommand{\nnmb}{\nonumber}
\newcommand{\del}{\partial}

\newcommand{\lrf}[2]{\left(\frac{#1}{#2}\right)}

\newcommand{\kev}{\, {\rm keV}}
\newcommand{\mev}{\, {\rm MeV}}
\newcommand{\gev}{\, {\rm GeV}}

\newcommand{\vmin}{v_{min}}
\newcommand{\vesc}{v_{esc}}
\newcommand{\erf}{\mathrm{erf}}

\begin{document}

\title{Neutrino Backgrounds in Future Liquid Noble Element\\
\vspace{0.35cm} Dark Matter Direct Detection Experiments}

\author{Andr\'ea~Gaspert}
\email{agaspert@stanford.edu}
\affiliation{Department of Physics, Stanford University, Stanford, CA 94305, United States}
\affiliation{TRIUMF, 4004 Wesbrook Mall, Vancouver, BC V6T 2A3, Canada}
\affiliation{D\'epartement de Physique, Universit\'e de Montr\'eal, C.P. 6128,
Succursale Centre-Ville, Montreal, Canada H3C 3J7}

\author{Pietro Giampa}
\email{pgiampa@snolab.ca}
\affiliation{SNOLAB,  1039 Regional Road 24, Lively,  ON  P3Y 1N2,  Canada}
\affiliation{TRIUMF, 4004 Wesbrook Mall, Vancouver, BC V6T 2A3, Canada}

\author{David~E.~Morrissey}
\email{dmorri@triumf.ca}
\affiliation{TRIUMF, 4004 Wesbrook Mall, Vancouver, BC V6T 2A3, Canada}

\date{\today}

\begin{abstract}
Experiments that use liquid noble gases as target materials, such as argon and xenon, play a significant role in direct detection searches for WIMP(-like) dark matter. As these experiments grow in size, they will soon encounter a new background to their dark matter discovery potential from neutrino scattering off nuclei and electrons in their targets. Therefore, a better understanding of this new source of background is crucial for future large-scale experiments such as ARGO and DARWIN. In this work, we study the impact of atmospheric neutrino flux uncertainties, electron recoil rejection efficiency, recoil energy sensitivity, and other related factors on the dark matter discovery reach. We also show that a significant improvement in sensitivity can potentially be obtained, at large exposures, by combining data from independent argon and xenon experiments.
\end{abstract}

\maketitle

\setcounter{page}{2}

%
\section{Introduction\label{sec:intro}}
Understanding the nature of dark matter remains one of the most significant open questions in fundamental science. Despite the strong evidence from astronomical measurements~\cite{Zwicky:1933gu,Rubin:1980zd,Kaiser:1992ps,Peebles:1991ch,Aghanim:2018eyx}, very little is known about what dark matter is made of. This puzzle has given rise to a global effort to develop highly sensitive experiments to discover and identify the nature of dark matter~\cite{Battaglieri:2017aum}. These experimental methods include direct detection~\cite{Goodman:1984dc,Drukier:1986tm,Lewin:1995rx}, indirect detection~\cite{Gunn:1978gr,Stecker:1978du,Stecker:1978du,Cirelli:2010xx,Slatyer:2015jla,Gaskins:2016cha}, and collider searches~\cite{Goodman:2010ku,Fox:2011pm,Abdallah:2015ter,Sirunyan:2017hci,Aaboud:2019yqu}.

The focus of this paper are direct detection experiments that look for local dark matter scattering on target nuclei in deep underground detectors~\cite{Goodman:1984dc,Drukier:1986tm,Lewin:1995rx}. Such experiments are particularly well-suited to finding weakly interacting massive particles~(WIMPs) and WIMP-like dark matter~\cite{Hut:1977zn,Lee:1977ua,Goldberg:1983nd,Jungman:1995df,Bertone:2004pz}. These are among the most promising and best-studied candidates for dark matter. Most notably, they can obtain the observed dark matter relic abundance during the evolution of the early universe through the mechanism of thermal freeze-out~\cite{Lee:1977ua}. New particles with masses near the weak scale are also predicted by most theories that address the electroweak hierarchy problem, and indeed many of them contain viable WIMP(-like) dark matter candidates~\cite{Feng:2010gw,Roszkowski:2017nbc}.

Current direct detection bounds on WIMP dark matter candidates with masses above a few GeV (with mainly spin-independent interactions with nuclei) are dominated by experiments using liquid noble gases as their target material, primarily xenon and argon. Existing experiments have already achieved exposures of multi-tonne years while controlling backgrounds, as demonstrated in Refs.~\cite{Lebedenko:2008gb, Akerib:2016vxi, Cui:2017nnn,Aprile:2018dbl, Agnes:2018fwg, Ajaj:2019imk,Wang:2020coa,PandaX-4T:2021bab}. In coming years, the LZ~\cite{Akerib:2018lyp}, PandaX-4T~\cite{Zhang:2018xdp}, and XENONnT~\cite{Aprile:2020vtw} are expected to achieve exposures near $20\,\text{t\,yr}$ in xenon, while DarkSide20k~\cite{Aalseth:2017fik} aims for a total exposure near $100\,\text{t\,yr}$ in argon. Beyond this, work is already under way on DARWIN based on xenon with an exposure goal of nearly $200\,\text{t\,yr}$~\cite{Aalbers:2016jon}, and ARGO based on argon reaching an exposure of $3000\,\text{t\,yr}$~\cite{Galbiati:2018}.

As such dark matter detectors grow in size and sensitivity, neutrinos from the sun~\cite{Bahcall:2004pz,Robertson:2012ib}, cosmic rays in the atmosphere~\cite{Barr:1989ru,Battistoni:2005pd,Honda:2006qj}, and diffuse supernovae~\cite{Hartmann:1997qe,Ando:2004hc,Beacom:2010kk} will become significant backgrounds to dark matter searches~\cite{Monroe:2007xp,Dodelson:2008yx,Strigari:2009bq,Billard:2013qya,Ruppin:2014bra,Gelmini:2018ogy,Newstead:2020fie,OHare:2020lva}. These neutrinos can scatter coherently with the target nucleus as a whole~\cite{Freedman:1973yd}, as recently measured for the first time (for reactor neutrinos) in CsI(Na)~\cite{Akimov:2017ade} and argon~\cite{Akimov:2020pdx}. In liquid noble element detectors this produces a nearly irreducible background to dark matter scattering on nuclei. Neutrinos can also scatter with electrons in the target material to produce a further background due to the finite ability of detectors to differentiate between nuclear and electronic interactions, especially close to the nuclear scattering detection energy threshold of a few keV~\cite{Ajaj:2019imk,Aalbers:2016jon}.

If the neutrino fluxes and energy spectra were known precisely, neutrino scattering would present a challenging statistical background, reducing the scaling of sensitivity with total exposure to a square root rather than linear~\cite{Billard:2013qya,Ruppin:2014bra}. However, the full story is more complicated. On the one hand, differences between the recoil energy spectra from dark matter scattering relative to neutrinos, due to a combination of their respective particle physics cross sections and fluxes, provide a small degree of distinction for the signal. Unfortunately, on the other hand, uncertainties in the neutrino fluxes induce a more difficult systematic background that is hard to overcome when they grow larger than the size of the dark matter signal itself. These uncertainties can ultimately lead to a so-called neutrino floor beyond which it is difficult to make progress in the direct detection approach~\cite{Billard:2013qya,Ruppin:2014bra}.

Future dark matter experiments based on xenon and argon, such as DARWIN~\cite{Aalbers:2016jon} and ARGO~\cite{Galbiati:2018}, are expected to reach nearly down to their respective neutrino floors throughout almost the entire range of WIMP(-like) dark matter masses they are designed to look for. Moreover, a broad range of WIMP(-like) dark matter theories predict nuclear scattering cross-sections near or below these detection-floors~\cite{Ellis:2005mb,Cheung:2012qy,Hill:2013hoa,Cahill-Rowley:2014boa,Bramante:2015una,Claude:2021sye}. To optimize the dark matter reach of these experiments, it is therefore crucial to understand the neutrino backgrounds and predict how they depend on the properties of the detectors such as total exposure, electron rejection efficiency, and recoil energy sensitivity. 

In this work, we investigate neutrino backgrounds to searches for WIMP(-like) dark matter through nuclear recoils in large-scale xenon and argon detectors. Our paper builds on previous work on neutrino backgrounds in three main ways~\cite{Monroe:2007xp,Dodelson:2008yx,Strigari:2009bq,Billard:2013qya,Ruppin:2014bra,Gelmini:2018ogy,Newstead:2020fie,OHare:2020lva}. First, we update detector energy sensitivity regions and resolutions for both future xenon and argon experiments, with a greater emphasis on argon relative to previous works. Second, we investigate the impact of electron recoil rejection on the sensitivity to dark matter subject to neutrino-electron scattering backgrounds. And third, we show that combining data from argon and xenon experiments can improve the sensitivity to dark matter beyond just an increase in statistics. Along the way, we also study the effects of detector location, energy detection regions of interest, and spectral shape uncertainties on dark matter detection.
 
The outline of this paper is as follows. We begin in Section~\ref{sec:scatt} by reviewing the calculations of the scattering rates for dark matter and background neutrinos in dark matter direct detection experiments. Next, in Section~\ref{sec:detectors} we present the motivating assumptions we make about the detection capabilities of future detectors as well the statistical methods we use in our analysis. Section~\ref{sec:nufloor} contains our main results on how the neutrino floor in xenon and argon is impacted by neutrino flux uncertainties, electron recoil rejection, and recoil energy sensitivity, as well as the effect of combining data from different future experiments. Our conclusions are presented in Section~\ref{sec:conc}. Some additional background material is summarized in Appendices~\ref{sec:appa}, \ref{sec:appb}.

\section{Dark Matter and Neutrino Scattering\label{sec:scatt}}
In this section, we review the calculations of dark matter and neutrino scattering rates. We also expand on previous treatments of neutrino scattering on electrons in detector targets.

\subsection{Dark Matter Scattering}\label{sec:dmscat}
The rate $\widetilde{R}_{\chi}^{(N)}$ of dark matter $\chi$ scattering off a nuclear species $N = (A, Z)$ per unit target mass is~\cite{Lewin:1995rx,Jungman:1995df}
\beq
\frac{d\widetilde{R}_\chi^{(N)}}{dE_R~} = n_N\lrf{\rho_{\chi}}{m_{\chi}}\,
\int_{v_{min}}\!\!d^3v\,v\,f_{lab}(\vec{v})\;\frac{d\sigma_N}{dE_R}\;
\label{eq:dmrate1}
\eeq
where $E_R$ is the nuclear recoil energy, $m_{\chi}$ is the dark matter mass, $\rho_{\chi}$ is the local dark matter energy density, $n_N$ is the number of $N$ nuclei per unit target mass, and $d\sigma_{N}/dE_R$ is the differential scattering cross-section. The integral in Eq.~\eqref{eq:dmrate1} runs over the dark matter velocity $\vec{v}$ in the laboratory frame, with $f_{lab}(\vec{v})$ being the lab-frame dark matter velocity distribution function. This integral is limited by $v > v_{min}$, the minimum dark matter velocity needed to produce a recoil of energy $E_R$. For elastic scattering, $v_{min}$ is given by
\beq
\vmin(E_R) = \sqrt{\frac{m_NE_R}{2\mu_N^2}} \ ,
\eeq
with $\mu_N = m_\chi m_N/(m_{\chi}+m_N)$ the dark matter-nucleus reduced mass. Note as well that the number of $N$ nuclei per unit detector mass $n_N$ can be expressed as
\beq
n_N = \frac{\mathcal{F}_N}{m_N} \ ,
\eeq
where $m_N$ is the mass of a single $N$ nucleus and $\mathcal{F}_N$ is the mass fraction of $N$ in the detector. 

In this work, we focus on the broad range of WIMP and WIMP-like dark matter theories in which the dominant nuclear scattering interaction is spin-independent and mediated by massive intermediate particles. In this case, the differential cross-section takes the form~\cite{Lewin:1995rx,Jungman:1995df}
\beq
\frac{d\sigma_{N}}{dE_R} = \frac{m_N}{2\mu_N^2v^2}\;\bar{\sigma}_N
\left|F_N(E_R)\right|^2 \ .
\label{eq:dsigder}
\eeq
Here $\bar{\sigma}_N$ depends on the nuclear target but is independent of $v$ and $E_R$, and $F_N(E_R)$ is a nuclear form factor that is approximated well by~\cite{Engel:1992bf}
\beq
F_N(E_R) = \frac{3\left[\sin(q\,r_N)-(q r_N)\cos(q r_N)\right]}{(qr_N)^3}\;
e^{-(qs)^2/2} \ ,
\label{eq:helm}
\eeq
for $s = 0.9\,\mathrm{fm}$,
\beq
q=\sqrt{2m_NE_R} \ ,
\eeq
and
\beq
r_N = \sqrt{ c^2 + (7\pi/3)\,\bar{a}^2 - 5 s^2} \ ,
\eeq
with $\bar{a} = 0.52\,\mathrm{fm}$ and $c = (1.23\,A^{1/3}-0.6)\,\mathrm{fm}$. To compare the sensitivities of dark matter detectors with different targets, it is also convenient to define a per-nucleon, spin-independent cross-section:
\beq
\sigma_{n} ~\equiv~ \frac{1}{A^2}\frac{\mu_n^2}{\mu_N^2}\,\bar{\sigma}_N \ ,
\eeq
where $\mu_n = m_{n}m_{\chi}/(m_n+m_{\chi})$. Given the dark matter cross-section form of Eq.~\eqref{eq:dsigder}, the scattering rate of Eq.~\eqref{eq:dmrate1} can be written in the form
\beq
\frac{d\widetilde{R}_\chi^{(N)}}{dE_R} = 
n_N\,m_N\,A^2\,\left|F_N(E_R)\right|^2\;\frac{\sigma_{n}}{2m_{\chi}\mu_p^2}\;
\rho_{\chi}\,\eta(v_{min})
\ .
\label{eq:dmrate2}
\eeq
The last term represents the integral over the dark matter halo and is given by
\beq
\eta(v_{min}) = \int_{v_{min}}\!\!d^3v\;\frac{f_{lab}(\vec{v})}{v} \ .
\label{eq:haloint}
\eeq
We evaluate the halo integral using the Standard Halo Model~(SHM) with the parameters $v_0=238\,\text{km/s}$, $v_E = 254\,\text{km/s}$, and $v_{esc} = 544\,\text{km/s}$, as well as $\rho_\chi = 0.3\,\gev/\text{cm}^3$. These are the recommended values from Ref.~\cite{Baxter:2021pqo} and they are in line with the values used in setting limits by recent direct dark matter search experiments~\cite{Akerib:2016vxi,Cui:2017nnn,Aprile:2018dbl}.  In evaluating Eq.~\eqref{eq:dmrate2}, we sum this expression over all isotopes of the target material weighted by their natural abundances.

Dark matter can also scatter off atomic electrons in the target material of the detectors. However, DM-electron scattering produces a much weaker signal than the nuclear recoil searches we focus on in this study for two reasons. First, the energy transfer from weak-scale dark matter to electrons is much less efficient when compared to scattering off nuclei, and very few electron scattering events produce visible energies above the nuclear recoil equivalent thresholds of $E_R\gtrsim 5\,\kev$ we focus on here~\cite{Dedes:2009bk,Kopp:2009et,Essig:2011nj}. Second, the searches we investigate reject electron recoils to a very high degree as a way to reduce backgrounds. For both reasons, we neglect dark matter-electron recoils in our analysis.

\subsection{Neutrino Scattering on Nuclei}\label{sec:nson}
Coherent elastic scattering of neutrinos on nuclei was observed recently for the first time in CsI(Na)~\cite{Akimov:2017ade} and Ar~\cite{Akimov:2020pdx} detectors, with rates that match Standard Model predictions. The same scattering processes from solar, supernova, and atmospheric neutrinos in dark matter detectors will become a significant background for future experiments as they become more sensitive. At leading order in the Standard Model, the differential cross-section for this process on a target nucleus $N = (A, Z)$ comes from $Z^0$ vector exchange and is given by~\cite{Freedman:1977xn}
\beq
\frac{d\sigma_{\nu N}(E_{\nu},E_R)}{dE_R} 
=  \frac{G_F^2 Q_W^2}{4\pi}\,m_N\,\left(1-\frac{m_NE_R}{2E_{\nu}^2}\right)
|F_N(E_R)|^2 \ ,
\label{eq:nunuc1}
\eeq
 where $E_\nu$ is the incident neutrino energy, $G_F$ is the Fermi constant, $F_N(E_R)$ is the Helm form factor defined in Eq.~\eqref{eq:helm}, and
\bea
Q_W = (A-Z) - Z(1-4s^2_W) \ ,
\eeq
 with $s^2_W = \sin^2\theta_W \simeq 0.23$~\cite{Zyla:2020zbs}. Since this process is mediated by $Z^0$ boson exchange, it is independent of neutrino flavour. It is only allowed kinematically for $E_{\nu} > E_{min}$ with
\beq
E_{min} = \sqrt{m_NE_R/2} \ .
\eeq
 Therefore, given the nuclear scattering cross-section, the rate per unit target mass $\widetilde{R}_{\nu}^{(N)}$ of neutrino scattering on the nuclear species, $N=(A, Z)$ is
\beq
\frac{d\widetilde{R}_{\nu}^{(N)}}{dE_R}
 = n_N\int_{E_{min}}\!\!dE_{\nu}\,
\sum_{j}\Phi_j(E_{\nu})\;\frac{d\sigma_{\nu N}}{dE_R} 
\equiv 
\sum_j\frac{d\widetilde{R}_{\nu\,j}^{(N)}}{dE_R}
\ ,
\label{eq:nunuc2}
\eeq
where $n_N$ is the number of $N$ nuclei per unit target mass, and the sum runs over all energy-differential neutrino fluxes $\Phi_j(E_{\nu})$ with $j=1,\ldots,n_{\nu}$ indexing the relevant sources. For each source, we have also implicitly summed over all neutrino flavours: $\Phi_j(E_{\nu}) = \sum_a\Phi_{j,a}(E_{\nu})$ for $a=e,\mu,\tau,\bar{e},\bar{\mu},\bar{\tau}$. In searches looking for heavier dark matter that induce nuclear recoils with $E_R \sim 10$--$100\,\kev$, the dominant nuclear recoil background comes from atmospheric neutrinos with energies in the range $E_{\nu} \sim 30$--$1000\,\mev$. Further details on neutrino sources relevant for dark matter searches are discussed in Sec.~\ref{sec:sources} below.

\subsection{Neutrino Scattering on Electrons}\label{sec:nsoe}
Neutrinos can also scatter elastically off of electrons. Such recoils in dark matter detectors can be misidentified as nuclear recoils. The probability for this to occur depends on the particle identification power of the experiment. Thus, these neutrino-induced electronic interactions can contribute to the background budget of dark matter searches based on nuclear scattering. The neutrino-electron scattering rate per unit target mass of neutral atoms containing the nucleus $N=(A, Z)$ from neutrino species $a=e,\mu,\tau,\bar{e},\bar{\mu},\bar{\tau}$ is
\beq
\frac{d\widetilde{R}_{\nu_a}^{(Ze)}}{dT} = 
n_N\,\int_{E_{min}}\!dE_{\nu}\;\sum_i\Phi_{i,a}(E_{\nu})\,\frac{d\sigma_a^{(Ze)}}{dT} \ ,
\eeq
 where $T$ is the energy transferred by the neutrino to the target and $d\sigma_a^{(Ze)}/dT$ is the total differential cross-section for electron scattering on the atomic target. 

To discuss the cross-section $d\sigma_a^{(Ze)}/dT$, it is convenient to begin with the cross-section for elastic neutrino scattering off a free electron, $\nu_a+e^-\to \nu_a+e^-$, given by ~\cite{Sarantakos:1982bp,Vogel:1989iv,Marciano:2003eq,Formaggio:2013kya}
\beq
\frac{d\sigma_a^{(e)}}{dT} = \frac{2\,G_F^2}{\pi}\,m_e
\left[
Q_-^2 
+ Q_+^2\Big(1-\frac{T}{E_{\nu}}\Big)^2
- Q_-\,Q_+\,\frac{m_eT}{E_{\nu}^2}
\right]\,\Theta(E_{\nu}-E_{min})
\eeq
where the parameters $Q_+$ and $Q_-$ are 
\beq
Q_- &=& \delta_{a\,e} - \frac{1}{2} + s^2_W\\
Q_+ &=& s^2_W
\eeq
and the minimum energy transfer $E_{min}$ is
\beq
E_{min} = \frac{1}{2}\left[T + \sqrt{T(T+2m_e)}\right] \ .
\eeq
In many analyses of neutrino backgrounds for dark matter detection, neutrino scattering off electrons is computed using the free-electron approximation~(FE) in which the atomic electrons of the target are treated as being unbound and at rest. The total neutrino cross-section off atoms of the nuclear species $N=(A, Z)$ in the FE approximation is
\beq
\frac{d\sigma_a^{(Ze,FE)}}{dT} = Z\,\frac{d\sigma_a^{(e)}}{dT} \ .
\label{eq:fea}
\eeq
The intuitive motivation underlying this approximation is that the energy transferred by the neutrino to the electron is relatively large compared to (most of) the atomic binding energies, and thus atomic effects should not be important. Indeed, for neutrino-induced electron recoils with $T = 1$--$100\,\kev$ relevant for weak-scale dark matter searches, the total scattering rate is dominated by solar neutrinos with energies between about $E_{\nu} \sim 0.1$--$1.0\,\mev$, with the largest contributions from the pp continuum and $^7$Be line sources. 

Going beyond the free-electron approximation, atomic effects on neutrino-electron scattering were studied in Refs.~\cite{Voloshin:2010vm,Kouzakov:2011vx,Kouzakov:2011ka,Kouzakov:2014lka,Chen:2013lba,Chen:2014ypv,Chen:2016eab,Hsieh:2019hug}. In Ref.~\cite{Voloshin:2010vm}, it was argued that atomic effects are small when $E_{\nu}\gg T,\,E_n$, where $E_n \lesssim \alpha^2Z^2m_e/2n^2$ are the relevant atomic binding energies.\footnote{This contrasts with non-relativistic DM scattering on atomic electrons, where atomic effects can be very significant\cite{Kopp:2009et,Essig:2011nj,Essig:2015cda}. For non-relativistic DM scattering on electrons with momentum transfer $q$ and energy transfer $T$ one has $T \ll q$, while for relativistic neutrinos $T \sim q$.} This work was expanded in Refs.~\cite{Kouzakov:2011vx,Kouzakov:2011ka,Kouzakov:2014lka}, which proposed a simple stepping approximation to capture the leading atomic effects in the regime $E_{\nu}\gg T,\,E_n$. This approximation extends the free-electron approximation formula of Eq.~\eqref{eq:fea} by replacing $Z$ with $Z_{eff}(T)$, given by
\beq
Z_{eff} = \sum_n\,\Theta(T-E_n) \ ,
\label{eq:zeff}
\eeq
where the sum runs over all relevant atomic levels $n$ with binding energies $E_n$. Careful \emph{ab initio} calculations of the neutrino-electron differential cross-section in atoms were performed in Refs.~\cite{Chen:2013lba,Chen:2014ypv} for germanium and in Ref.~\cite{Chen:2016eab} for xenon. When considering pp and $^7$Be solar neutrinos scattering on atomic electrons in xenon, Ref.~\cite{Chen:2016eab} found that the full result is similar to the stepping approximation but smaller by about 25\% up to $T\sim 10\,\kev$ and then falls off more quickly at higher energies. In the analysis to follow, we apply the results of Ref.~\cite{Chen:2016eab} for neutrino-electron scattering on xenon but use the stepping approximation for neutrino-electron scattering on argon where it is expected to be a better approximation due to its lower binding energies ($E_n \lesssim 3.2\,\kev$ in Ar versus $E_n \lesssim 35\,\kev$ in Xe). We also assume that the energy transfer $T$ produces visible energy in the detector with the same efficiency as free-electron recoils.

\subsection{Neutrino Fluxes and Scattering Rates\label{sec:sources}}

\begin{figure}
  \centering
    \includegraphics[width = 0.7\textwidth]{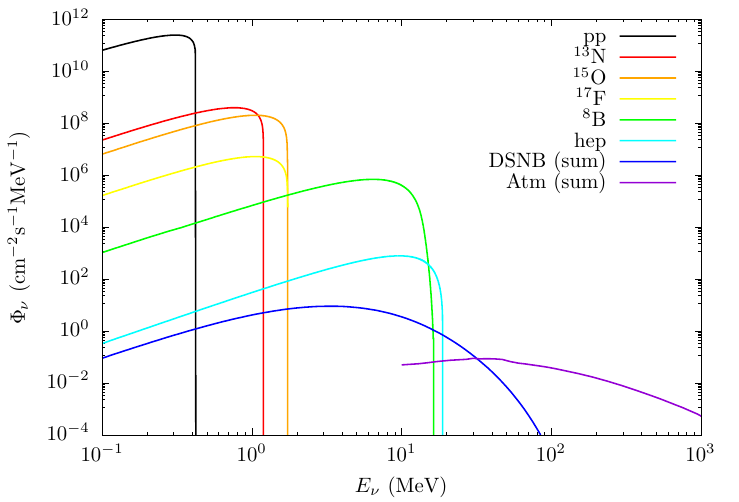}
  \caption{Neutrino flux spectra relevant for dark matter   direct detection with overall normalizations and source models as listed in Table~\ref{tab:flux}.\\
}
  \label{fig:flux}
\end{figure}

The scattering kinematics discussed above imply that the neutrino backgrounds relevant for the direct detection of weak-scale dark matter correspond to neutrino energies $E_\nu \sim 1$--$1000\,\mev$, for neutrino-nucleus scattering and $E_{\nu} \sim 0.1$--$1\,\mev$ for neutrino-electron scattering. In these energy ranges, the most important neutrino flux contributions come from atmospheric neutrinos created by cosmic ray showers~\cite{Barr:1989ru,Battistoni:2005pd,Honda:2006qj,Honda:2011nf}, solar neutrinos from nuclear reactions in the Sun~\cite{Bahcall:2004pz,Robertson:2012ib,Bahcall,Bahcall:2004fg,Bahcall:2004mq}, and the diffuse supernova neutrino background (DSNB)~\cite{Hartmann:1997qe,Ando:2004hc,Beacom:2010kk}. The neutrino flux spectra for the various sources are shown in Fig.~\ref{fig:flux} and are based on the source models listed in Table~\ref{tab:flux}. Each independent flux source $j$ can be written in the form
\beq
\Phi_j(E_{\nu}) = \phi_j\,f_j(E_{\nu}) \ , 
\label{eq:flux}
\eeq
where $\phi_j$ is the total flux and $f_j(E_{\nu})$ is an energy distribution normalized to unity. For reference, we list in Table~\ref{tab:flux} the flux normalizations used in this analysis together with their estimated uncertainties. 

Atmospheric neutrinos are the most challenging background to dark matter direct detection for masses above about $m_{\chi} \gtrsim 10\,\gev$: they are the dominant neutrino source at energies above $E_{\nu}\gtrsim 30\,\mev$ and the nuclear recoil energy spectrum they induce aligns closely with that of dark matter. Calculations of neutrino fluxes in the relevant energy range include Refs.~\cite{Barr:1989ru,Battistoni:2005pd,Honda:2006qj,Honda:2011nf,Barr:2004br,Barr:2006it,Battistoni:2002ew,Honda:2015fha}. These calculations agree reasonably well with each other and with direct measurements of atmospheric neutrino fluxes by the Super-Kamiokande experiment down to energies of a few hundred MeV~\cite{Richard:2015aua}. The estimated uncertainties on these calculations over the relevant energy range are roughly $20\%$~\cite{Honda:2011nf}, but with much smaller fractional uncertainties of about $5\%$ in the flux ratios $\nu_{e}/\nu_{\mu}$ and $\bar{\nu}/\nu$~\cite{Battistoni:2005pd,Honda:2011nf,Barr:2006it}. In this work, we use the neutrino fluxes calculated in Ref.~\cite{Battistoni:2005pd} for the simple reason that they are the most recent results that extend to the lower energies needed for this analysis, down to $E_{\nu} = 10\,\mev$. The fluxes in these works are calculated at the mean of the solar cycle for the Laboratori Nazionali del Gran Sasso~(LNGS) and Super Kamiokande sites, and we use the LNGS result throughout this work.\footnote{Relative to LNGS, the atmospheric fluxes computed in Ref.~\cite{Honda:2011nf} for SNOLAB are larger by about $30\%$.} Since the flux uncertainties between different neutrino flavours are strongly correlated, and neutrino-nucleus scattering is the same for all flavours, we work exclusively with the total atmospheric flux summed over flavours with a fractional uncertainty of $20\%$. 

Among the solar neutrino sources, the most important contributions to neutrino-nucleus scattering in the detectors we study come from the $^8$B and hep fluxes. For neutrino-electron scattering, instead, the most dominant contributions come from the $pp$ and $^7$Be$~(862\,\kev)$ fluxes due to the differences in kinematics between electron and nuclear scattering. To model solar neutrino fluxes, we use the flux shapes $f_j(E_\nu)$ collected in Ref.~\cite{Bahcall:2004pz,bahcallshape} combined with updated estimates for the flux normalizations $\phi_j$. These are obtained from recent calculations of solar neutrino fluxes based on solar models as well as neutrino data when available. Our baseline is the set of solar model predictions from Ref.~\cite{Vinyoles:2016djt} for the high metallicity~(HZ) scenario of Ref.~\cite{Grevesse:1998bj,Asplund:2009fu} as updated in Ref.~\cite{Serenelli:2011py}. This model appears to give a better agreement with observations than other proposals~\cite{Vinyoles:2016djt,Wurm:2017cmm}. For the $^8$B flux, we use the determination of Ref.~\cite{Bergstrom:2016cbh} based on data from SNO~\cite{Aharmim:2011vm} and SuperK~\cite{Abe:2016nxk} combined with fits to neutrino oscillation data. The $^7$Be~($862\,\kev)$ line flux adopted in this work comes from the recent measurement by Borexino~\cite{Agostini:2020mfq}. We use the $^7$Be branching ratios from Ref.~\cite{Tilley:2002vg} to fix the flux of the related $^7$Be~(384\,keV) line in terms of the $^7$Be~(862\,keV) flux. The rate of neutrino-electron scattering relevant for dark matter searches also depends on the electron-neutrino fraction $\mathcal{F}_e$ of the solar neutrino flux from $pp$ and $^7$Be~($862\,\kev$); for both channels, we use the predicted and approximately constant MSW LMA~\cite{Wolfenstein:1977ue,Mikheev:1986gs,Mikheev:1986wj} value $\mathcal{F}_e = 0.55\,(1\pm 0.02)$ based on recent determinations of the neutrino oscillation parameters collected in Ref.~\cite{Zyla:2020zbs}. 

The collected emission from many supernovae over the history of the cosmos forms a diffuse supernova neutrino background~(DSNB). This source is the dominant flux in the energy range $E_{\nu} \sim 20$--$40\,\mev$. To model the DSNB fluxes of the $e$, $\bar{e}$, and $x = \mu,\,\bar{\mu},\,\tau,\,\bar{\tau}$ flavours (at production), we apply the methods of Ref.~\cite{Beacom:2010kk}. In doing so, we follow Ref.~\cite{Jeong:2018yts} and use the parametrization of the star formation rate from Ref.~\cite{Yuksel:2008cu} and connect it to the cosmic supernova rate based on Ref.~\cite{Horiuchi:2011zz}. The neutrino fluxes per supernova are based on Ref.~\cite{Beacom:2010kk} and use a simple Fermi-Dirac distribution with $E_{\nu}^{tot} = 3\times 10^{53}\,\mathrm{erg}$, and effective temperatures $T_a = 4,\,5,\,8~\mev$ for $a=e,\,\bar{e},\,x$. These estimates have a number of uncertainties associated with them, and in our analysis we apply an overall 50\% uncertainty on the summed DSNB rate. 

\begin{figure}[ttt]
  \begin{center}
    \includegraphics[width = 0.47\textwidth]{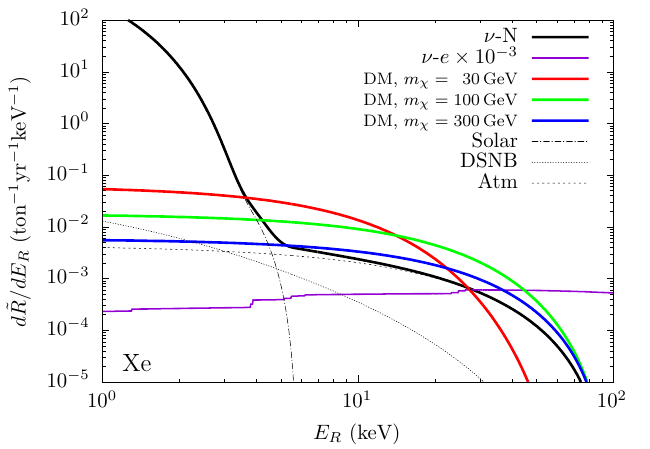}
    \includegraphics[width = 0.47\textwidth]{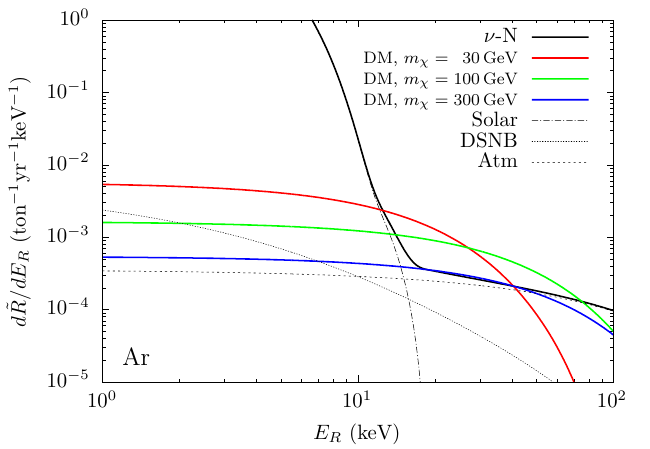}
  \end{center}
  \caption{Differential scattering rates per unit target mass in xenon~(left) and argon~(right) as a function of reconstructed nuclear recoil energy $E_{R}$ for dark matter-nucleus, neutrino-nucleus~($\nu$-N), and neutrino-electron~($\nu$-e) scattering. All dark matter curves assume $\sigma_n = 10^{-48}\,\text{cm}^2$. The $\nu$-e spectrum in xenon is scaled by a factor of $10^{-3}$ corresponding to a typical electron rejection power in xenon detectors.
\label{fig:spectra}}
\end{figure}

In Fig.~\ref{fig:spectra} we show the total differential scattering rates per unit target mass as a function of recoil energy $E_R$ for neutrino-nucleus scattering in xenon~(left) and argon~(right). We also show the corresponding rates for dark matter scattering with cross-section $\sigma_n=10^{-48}\,\text{cm}^2$ and masses $m_{\chi} = 30,\,100,\,300\,\gev$. For xenon, we also show the rate for neutrino-electron scattering in terms of the reconstructed nuclear-recoil equivalent energy and scaled by a factor of $10^{-3}$, which is on the order of the expected electron rejection power of future xenon detectors. No such line is shown for argon since future detectors are expected to be able to reject electron recoils to a very high degree. The dashed, dotted, and dot-dashed lines indicate the atmospheric~(Atm), supernova~(DSNB), and solar neutrino flux contributions to the neutrino-nucleus scattering rate. We see that the atmospheric flux is the dominant neutrino-nucleus background for $E_R \gtrsim 5\,\kev$ in xenon and $E_R\gtrsim 20\,\kev$ in argon. The recoil energy spectrum from atmospheric neutrinos is also similar to that from heavier $m_{\chi}\gtrsim 100\,\gev$ dark matter in xenon, while it is somewhat more distinct in argon. Neutrino-electron scattering can become important in xenon at higher recoil energies, even with strong electron recoil rejection, but its spectral shape is much flatter than for dark matter at these energies.

\begin{table}[ttt]
  \centering
  \begin{tabular}{c|c|c|c}
    Source & Flux ($\phi_j/\mathrm{cm}^{-2}\mathrm{s}^{-1}$) & Uncertainty\,(\%) & Ref. \\ \hline
    pp  & $5.98 \times 10^{10}$   & 0.6  & \cite{Vinyoles:2016djt}  \\\hline 
    \textsuperscript{7}Be~(862\,keV)  & $4.99 \times 10^{9}$  & 3.8& \cite{Agostini:2020mfq}  \\ \hline
    pep                    & $1.45 \times 10^{8}$ & 0.9 & \cite{Vinyoles:2016djt}    \\ \hline
    \textsuperscript{13}N   & $2.78 \times 10^{8}$ & 15 & \cite{Vinyoles:2016djt}    \\ \hline
    \textsuperscript{15}O  & $2.05 \times 10^{8}$ & 17 & \cite{Vinyoles:2016djt}    \\ \hline
    \textsuperscript{17}F  & $5.29 \times 10^{6}$ & 20 & \cite{Vinyoles:2016djt}    \\ \hline
    \textsuperscript{8}B  & $5.16 \times 10^{6}$ & 2.5 & \cite{Vinyoles:2016djt}    \\ \hline
    hep   & $7.98  \times 10^{3}$ & 30   & \cite{Vinyoles:2016djt}    \\ \hline
  Atm  & -- & 20 & \cite{Battistoni:2005pd} \\ \hline 
  DSNB  & -- & 50 & \cite{Keil:2002in} 
  \end{tabular}
  \caption{Overall normalizations $\phi_j$ of the neutrino fluxes included in our analysis together with the estimated fractional uncertainties we use for them. The solar fluxes from Ref.~\cite{Vinyoles:2016djt} are for the high metallicity~(HZ) scenario.}
  \label{tab:flux}
\end{table}

%
\section{Detecting and Characterizing Scattering Signals\label{sec:detectors}}
The signals induced by dark matter or neutrino scattering are shaped by the detectors used to measure them. In this section, we discuss the expected properties of future liquid noble element-based detectors and specify the assumptions we make to analyze their sensitivity. We also describe the statistical procedure used to test the discovery power of these experimental approaches.

\subsection{Properties of Liquid Noble Element Detectors}
Planned liquid noble element detectors will search for WIMP-like dark matter with mass above $m_{\chi}\gtrsim 10\,\gev$ primarily through nuclear recoils~(NR). Many of the backgrounds to these searches come from electron recoils~(ER). As such, these detectors are designed to have large acceptances for nuclear recoils together with the ability to reject the vast majority of electron recoils.  

Future large-scale xenon detectors are expected to have a dual-phase format similar to that used in ZEPLIN~\cite{Lebedenko:2008gb}, XENON~\cite{Aprile:2018dbl,Aprile:2010bt,Aprile:2017aty}, LUX~\cite{Akerib:2016vxi,Akerib:2012ys}, and Panda-X~\cite{Cui:2017nnn,Wang:2020coa,Cao:2014jsa}. Scattering in the liquid bulk of the detector leads to scintillation, ionization, and heat. The primary scintillation is detected directly, while the ionization electrons are drifted through an electric field to a cathode at the liquid-gas interface region where they create secondary photons. The energy deposited by scattering events can thus be characterized by the number of primary photons S1~(scintillation) together with the number of secondary photons S2~(ionization). Combining S1 and S2, with appropriate weights after correcting for the position, gives the total energy deposited by the event. Finally, the ratio S2/S1 is used to discriminate between nuclear and electron recoils. 

By extrapolating from the current capabilities of two-phase xenon detectors, future large-scale detectors are expected to have excellent sensitivity to nuclear recoils together with strong electron rejection power for recoil energies above about $E_R \gtrsim 5\,\kev$. Radioactive backgrounds to xenon dark matter searches tend to rise at higher recoil energies; to avoid the worst of them, the search region for dark matter scattering is typically limited from above to $E_R \lesssim 35\,\kev$. In the following analysis, we assume a region of interest~(ROI) for weak-scale dark matter searches in xenon experiments of $E_R \in [5,\,35]\,\kev$, in line with the projections of Refs.~\cite{Akerib:2018lyp,Zhang:2018xdp,Aprile:2020vtw,Aalbers:2016jon}. Within this ROI, cutting on the S2/S1 ratio is expected to allow the rejection of electron recoils by a factor of $10^{-4}$--$10^{-3}$ at the cost of reducing the acceptance for nuclear recoils to about $30$--$50\%$~\cite{Aalbers:2016jon}. For our analysis, we assume constant electron rejection factors over the entire ROI energy range between $\varepsilon_e = 2\times 10^{-4}$--$2\times 10^{-3}$ together with a realistic but optimistic nuclear recoil acceptance of $\varepsilon_n = 0.5$. These properties are summarized in Table~\ref{tab:detector}. 

Planned large-scale argon detectors~\cite{Aalseth:2017fik,Galbiati:2018} may rely on either a single-phase technique like DEAP-3600~\cite{Ajaj:2019imk,Amaudruz:2017ibl} or a dual-phase format such as DarkSide~\cite{Agnes:2018fwg,Aalseth:2017fik}. The scintillation yield of primary photons in argon is smaller than in xenon (Xe $\sim 60\,\gamma/\kev$ and Ar $\sim 40\,\gamma/\kev$~\cite{Chepel:2012sj, Szydagis:2011tk}), and this tends to translate into higher energy thresholds for NR when relying on primary scintillation. Based on the recent performance of DEAP-3600~\cite{Ajaj:2019imk}, we assume a nuclear recoil energy ROI in argon of $E_R \in [55,100]\,\kev$, although we also study the impact of reducing the lower energy threshold. A crucial feature of argon detectors is the ability to use pulse-shape discrimination to distinguish between nuclear and electronic interactions~\cite{Boulay:2006mb,Amaudruz:2016qqa,Adhikari:2020zyy}. Based on the properties of current argon dark matter detectors, we consider flat electron rejection factors in the range $\varepsilon_e = 10^{-9}$--$10^{-7}$ along with a nuclear recoil acceptance over the ROI of $\varepsilon_n = 0.9$~\cite{Ajaj:2019imk,Aalseth:2017fik,Adhikari:2021wrm}. Similar to xenon, the adopted argon detector parameters are reported in Table~\ref{tab:detector}.

In both xenon and argon, electron recoils create more visible energy (and less energy lost to heat) than nuclear recoils for a given recoil energy~\cite{Doke:2002oab}. As a result, an electron recoil that is misidentified as a nuclear recoil with energy $E_R$ corresponds to a smaller electron recoil energy $T$. The average relation between these quantities is defined as the quenching factor $q_{eff}$:
\beq
T = q_{eff}(E_R)\,E_R \ .
\eeq
This factor has been estimated using phenomenological models in Refs.~\cite{Lindhard:1963,Mei:2007jn,Bezrukov:2010qa,Sorensen:2011bd}, and investigated in direct calibration measurements in both argon~\cite{Brunetti:2004cf,Cao:2014gns,Agnes:2018mvl,Washimi:2018kgn} and xenon~\cite{Manzur:2009hp,Aprile:2009dv,Plante:2011hw}. A selection of these experimental results have been used by the NEST collaboration to make a global fit for $q_{eff}(E_R)$~\cite{Szydagis:2011tk,Szydagis:2021hfh}. In this work, we follow NEST and others and adopt a data-motivated model with the form
\beq
q_{eff}(E_R) = A_q\lrf{E_R}{\kev}^{\!B_q} \ ,
\label{eq:qmodel}
\eeq
where $A_q$ and $B_q$ are dimensionless coefficients. These parameters can depend on the strength of an applied electric field, which is relevant for dual-phase detectors. For xenon, the NEST collaboration quotes $A_q = 0.151\pm 0.027$ and $B_q=0.1\pm 0.05$ at zero applied electric field~\cite{Szydagis:2021hfh} while the recent LUX analysis finds $A_q=0.173$ and $B_q=0.05$ over a range of applied fields between about $30$--$600$~V/cm~\cite{Akerib:2020lkv}. In the xenon analysis to follow we fix $A_q=0.16$ and $B_q=0.08$, assuming an E-field strength between 50--200~V/cm. For argon, in zero applied field, we set $A_q=0.185$ and $B_q=0.12$ which is consistent with the NEST estimate of $A_q=0.19\pm 0.01$ (for $W=15.3\,\mathrm{eV}$) and $B_q=0.101\pm0.025$~\cite{Szydagis:2021hfh} as well as recent measurements by SCENE~\cite{Cao:2014gns} and ARIS~\cite{Agnes:2018mvl}.

Realistic detectors also have a finite recoil energy resolution. In modern xenon and argon detectors, most of the spread $\sigma_R$ in the reconstructed recoil energy, $E_R$, comes from stochastic fluctuations in the amounts of light and charge produced in individual scattering events~\cite{Sorensen:2011bd,Schumann:2015cpa}. To account for this factor in future xenon detectors, we assume a resolution similar to that achieved in the LUX~\cite{Akerib:2016qlr,Akerib:2016mzi} and XENON1T~\cite{Aprile:2020yad} experiments based on combined $S1$ and $S2$ energy determination. These experiments report an energy resolution for higher-energy electron recoils of $\sigma/T \simeq 0.32/\sqrt{T/\kev}$. We extrapolate this resolution to nuclear recoils by including the quenching factor discussed above,
\beq
\frac{\sigma_R}{E_R} = \frac{0.32}{\sqrt{q_{eff}(E_R)\,(E_R/\kev)}}
= {0.80}\lrf{E_R}{\kev}^{\!-0.54} \ .
\label{eq:xeres}
\eeq
The recent NEST determination of Ref.~\cite{Szydagis:2021hfh} argues that at electron energies below $T \sim 5\,\kev$, the resolution $\sigma/E \simeq 0.14$ saturates at a constant value. However, other interpretations exist for the resolution of xenon at these lower energies, and we apply the form of Eq.~\eqref{eq:xeres} at all energies, consistent with Ref.~\cite{Akerib:2016mzi}. For our projections in argon, we assume a nuclear recoil energy resolution equal to that achieved by the DEAP-3600 experiment~\cite{Ajaj:2019imk}, which over the energy ROI is fit accurately by the following parameterization:
\beq
\frac{\sigma_R}{E_R} = {1.09}\lrf{E_R}{\kev}^{\!-0.55} \ .
\label{eq:arres}
\eeq

To account for detector energy resolution, we use a binned statistical analysis (described in detail below) with recoil energy bins constrained to be larger than the local energy resolution. With an energy resolution of the general form $\sigma_R/E_R = A_r(E_R/\kev)^{-B_r}$, as we find for both argon and xenon, the maximum number of bins satisfying this constraint is
\beq
N_b = \lfloor {\mathcal{W}}/{(A_r\,B_r)}\rfloor \ ,
\eeq
where $\lfloor x\rfloor$ denotes the floor function and 
\beq
\mathcal{W} = \lrf{E_R^{max}}{\kev}^{B_r}-\lrf{E_R^{min}}{\kev}^{B_r} \ .
\eeq
For $i=1,2,\ldots N_b$, the $i$-th recoil energy bin then covers the 
range $E_R \in [E_{i-1},E_i]$ with
\beq
\lrf{E_i}{\kev} = \left[\lrf{E_R^{min}}{\kev}^{B_r}
+ \frac{i}{N_b}\mathcal{W}\right]^{1/B_r} \ . 
\eeq
This approach is expected to give a good approximation of the true energy resolution as long as the signal and background vary reasonably slowly from bin to bin, which we find to be the case in the examples studied here. 

In addition to events from dark matter and neutrinos, experiments also detect background events from various sources. Projections for future experiments suggest that these backgrounds can be reduced to be much less important than neutrino-induced backgrounds over the dark matter search ROIs in argon and xenon detectors. In both detector materials, the dominant nuclear recoil background comes from neutrons produced by radioactive decays; these can be mitigated effectively by the use of radiopure detector materials and active veto systems~\cite{Schumann:2015cpa,Aalbers:2016jon,Aalseth:2017fik,Galbiati:2018,Billard:2021uyg}. Electron recoil backgrounds arise primarily from nuclear decays of noble element contaminants in the detector material. They are expected to be subleading relative to electron recoils from solar neutrinos in xenon~\cite{Schumann:2015cpa}, while they are reduced to low levels by purification and electron recoil rejection in argon~\cite{Galbiati:2018}. Based on these considerations, we only consider backgrounds from neutrinos in our analysis. 

Putting these detector considerations together, the mean detection rate per unit target mass for events identified as nuclear recoils  with energy $E_R$ is
\beq
\frac{d\widetilde{R}^{(N)}}{dE_{R}} =
\varepsilon_N(E_R)\left(
\frac{d\widetilde{R}_{\chi}^{(N)}}{dE_R}
+ \frac{d\widetilde{R}_{\nu}^{(N)}}{dE_R}\right)
+ \left.\varepsilon_e(E_{R})\,\frac{dT}{dE_R}\,
\frac{d\widetilde{R}_{\nu}^{(Ze)}}{dT}\right|_{T=q_{eff}E_R} \ .
\eeq
 All detector parameters relevant for evaluating this expression are collected in Table.~\ref{tab:detector}. For each bin $i=1,2,\ldots N_b$, the expected number $N^i$ of events is then
\beq
N^i = M\,T\, \int_{E_{i-1}}^{E_i}\!\!dE_R\,\frac{d\widetilde{R}^{(N)}}{dE_{R}} \ ,
\eeq
 where $M\,T$ is exposure, equal to the total detector mass times the  observation time.

\begin{table}[ttt]
  \centering
  \begin{tabular}{ccc}
    \hline\hline
    & \textbf{Xenon \cite{Aalbers:2016jon}} & \textbf{Argon \cite{Aalseth:2017fik}} \\ \hline\hline
    \textbf{A, Z} & 131.293 u, 54 & 39.948 u, 18  \\ \hline
    \textbf{Energy ROI: $E_R^{min}$, $E_R^{max}~(\kev)$} & 5, 35&55, 100 \\ \hline
    \textbf{Nuclear Recoil Acceptance~($\varepsilon_e$)} & 0.50 & 0.90 \\ \hline
    \textbf{Electron Recoil Rejection~($\varepsilon_e$)} & ~~~$2\times 10^{-4}$ -- $2\times 10^{-3}$~~~ &~~~ $10^{-9}$ -- $10^{-7}$ \\ \hline
    \textbf{Quenching Model: $A_q$, $B_q$}~(Eq.~\eqref{eq:qmodel})&0.16, 0.08& 0.185, 0.12\\ \hline
    \textbf{Resolution Model: $A_r$, $B_r$}~(Eqs.~(\ref{eq:xeres},\ref{eq:arres}) &0.80, 0.54& 1.09, 0.55 \\ \hline   \hline
  \end{tabular}
  \caption{Considered properties for next-generation noble liquid dark matter detectors.}
\label{tab:detector}
\end{table}

\subsection{Statistical Methods}\label{sec:statmethod}

Following Refs.~\cite{Billard:2013qya,Ruppin:2014bra,Billard:2011zj,Billard:2013gfa}, we use the profile likelihood method~\cite{Cowan:2010js} to compute the dark matter discovery limit for a given set of experimental parameters, corresponding to the smallest dark matter cross-section for which the experiment can exclude the background-only hypothesis with a 3$\sigma$ significance at least 90\% of the time. The test statistic for our analysis is a ratio of likelihoods that compares the background-only hypothesis to the signal hypothesis. As discussed above, we assume binned data organized in recoil energy bins $i=1,2,\ldots,\,N_b$. The expected number of events in bin $i$ can thus be written in the form
\beq
\langle N^i\rangle = 
\xi\,s^i + b^i(\vec{\theta}) \ ,
\eeq
where $\xi = \sigma/\sigma_0$ is the $\chi$-nucleon cross-section relative to some reference value and $\vec{\theta}$ are a set of nuisance parameters that characterize the neutrino backgrounds. For a given set of data $\{N^i\}$, the likelihood function is taken to be
\beq
\mathcal{L}(\xi,\vec{\theta}) = \prod_{i=1}^{N_b}
\mathcal{P}(N^i;\langle N^i\rangle)
\times \mathcal{L}_b(\vec{\theta}) \ ,
\label{eq:lhood}
\eeq
 where $\mathcal{P}(n,\lambda) = \lambda^ne^{-\lambda}/n!$ is the Poisson distribution and $\mathcal{L}_b(\vec{\theta})$ is a likelihood function for the background (and possibly signal) parameters. The test statistic for our analysis is then
\beq
q_0 = -2\,\ln\left[
\mathcal{L}(\widetilde{\xi}=0,\widetilde{\vec{\theta}})
\Big/
\mathcal{L}(\hat{\xi},\hat{\vec{\theta}})
\right]\,\Theta(\hat{\xi}) \ .
\label{eq:q0}
\eeq
 Here, $\hat{\xi}$ and $\hat{\vec{\theta}}$ are the parameters that maximize the unconstrained likelihood of Eq.~\eqref{eq:lhood} for the set of data $\{N^i\}$, and $\widetilde{\vec{\theta}}$ are the values that maximize it for the background-only hypothesis. Larger values of $q_0$ correspond to poorer fits for the background hypothesis to the data.

It is now straightforward to formulate the dark matter discovery limit in a precise way. We follow Refs.~\cite{Billard:2013qya,Ruppin:2014bra,Billard:2011zj,Billard:2013gfa} and say that an experiment can discover a given dark matter scenario with cross-section strength $\xi$ if it is able to exclude the background hypothesis with $p$-value below $p < p_0 = 0.0013$ (corresponding to an exclusion of at least $3\sigma$ for a Gaussian distribution) at least a fraction $\mathcal{F} = 0.90$ of the time. Let us define $q_0^{lim}$ by the relation
\beq
\mathcal{F} = \int_{q_0^{lim}}^{\infty}\!dq_0\;f(q_0|\xi) \ .
\label{eq:ffrac}
\eeq
The condition to exclude the background hypothesis is then
\beq
p_0 > \int_{q_0^{lim}}^{\infty}\!dq_0\;f(q_0|0) \ .
\label{eq:pvalue}
\eeq

To evaluate the dark matter discovery condition of Eq.~\eqref{eq:pvalue} for a given scenario, we make use of a key feature of the test statistic of Eq.~\eqref{eq:q0}: its distribution has a simple asymptotic form for $N_{tot} = \sum_i\langle{N}^i\rangle \gg 1$. As discussed further below, this condition is found to hold for the scenarios of interest in this work for exposures $M\,T \gtrsim 10\,\text{t\,yr}$. By applying Wilks' theorem~\cite{Wilks:1938dza,Wald:1942}, it was shown in Ref.~\cite{Cowan:2010js} that the probability distribution for $q_0$ when the underlying data has value $\xi$ is
\beq
f(q_0|\xi) = \left( 1-\Phi(\sqrt{q_{0,A}})\right)\,\delta(q_0)
+ \frac{1}{2\sqrt{2\pi q_0}}\,
\exp\left[-\frac{1}{2}\left(\sqrt{q_0}-\sqrt{q_{0,A}}\right)^2\right] \ ,
\label{eq:wilks}
\eeq
with
\beq
\Phi(z) = \frac{1}{\sqrt{2\pi}}\int_{-\infty}^z\!dx\;e^{-x^2/2} \ ,
\eeq
and $q_{0,A}$ being the value of the test statistic evaluated on the \emph{Asimov data set} with $N^i = \langle N^i\rangle$. Note that for this specific dataset, $\hat{\xi} = \xi$ and $\hat{\vec{\theta}} = \vec{\theta}$ in general, and $q_{0,A}\to 0$ for $\xi \to 0$. 

Applying the asymptotic form of  Eq.~\eqref{eq:wilks} to Eqs.\,(\ref{eq:ffrac},\ref{eq:pvalue}), which hold in general, the condition of Eq.~\eqref{eq:pvalue} simplifies enormously to
\beq
q_{0,A} > \left[\Phi^{-1}(1-p_0) - \Phi^{-1}(1-\mathcal{F})\right]^2 \ . 
\label{eq:qzero}
\eeq
For $p < p_0 = 0.0013$ we have $\Phi^{-1}(1-p_0) = 3$ and $\mathcal{F} = 0.90$, and this reduces to $q_{0,A} > 18.34$. The \emph{dark matter discovery limit} is then the smallest value of $\xi$ for which $q_{0,A} > 18.34$. 

In the analysis to follow, we apply this asymptotic relation to estimate how the dark matter discovery limit varies with the size and properties of hypothetical future experiments. The approach requires only a single minimization over the space of $\vec{\theta}$ per scenario. We do this minimization using a global pre-conditioning step followed by a local minimization based on the MMA algorithm~\cite{Svanberg_aclass} as implemented in the NLopt optimization package~\cite{nlopt}. In contrast, a full treatment that does not use the asymptotic forms would require generating multiple sets of pseudo-data, each needing two minimizations, to estimate the distributions $f(q_0|\xi)$ and $f(q_0|0)$. Applied to simple examples, Refs.~\cite{Cowan:2010js,Burns:2011xf,Bhattiprolu:2020mwi,Xia:2021ola, Basso:2021dwp} find that convergence is reasonably good for $N_{tot} \gtrsim 10$. This condition is met for the main scenarios of interest in this work with exposure $M\,T \gtrsim 10\,\text{t\,yr}$. We have also verified for several specific scenarios that the discovery limits derived with the asymptotic results based on the Asimov data set agree very well with the limits obtained with simulated pseudo-datasets.

\subsection{Characterizing Uncertainties}

To compute the likelihoods needed to obtain dark matter discovery limits, it is necessary to specify the background-parameter likelihood function $\mathcal{L}_b(\vec{\theta})$ in Eq.~\eqref{eq:lhood}. In this work, we focus on neutrinos as the primary background to dark matter. Other background sources can be mitigated such that they are much less important over the dark matter regions of interest that we study~\cite{Schumann:2015cpa,Aalbers:2016jon,Aalseth:2017fik}. The main background uncertainties are therefore the neutrino fluxes from the $n_{\nu}$ sources considered. For these, we mostly fix the shapes of the flux energy spectra $f_j(E_{\nu})$ and allow for variations only in the overall flux normalizations $\phi_j$, as defined in Eq.~\eqref{eq:flux}. It is therefore convenient to describe flux uncertainties with the variables
\beq
\theta_j \equiv \frac{\phi_j-\bar{\phi}_j}{\bar{\phi}_j} \ ,
\eeq
 with $j=1,\ldots n_{\nu}$ and $\bar{\phi}_j$ being the central flux values listed in Table~\ref{tab:flux}. In terms of these variables, we take the background likelihood function to be a product of independent Gaussians for each source,
\beq
\mathcal{L}_b(\vec{\theta}) = \prod_{j=1}^{n_{\nu}}
\frac{1}{\sqrt{2\pi}\Delta\theta_j}\,
e^{-({\theta_j}/{\Delta\theta_j})^2/2}
\ ,
\label{eq:bglhood}
\eeq
 where $\Delta\theta_j$ coincide with the fractional flux uncertainties listed in Table~\ref{tab:flux}.

\section{Dark Matter Detection and the Neutrino Floor\label{sec:nufloor}}

\begin{figure}[ttt]
  \centering
  \includegraphics[width=.47\textwidth]{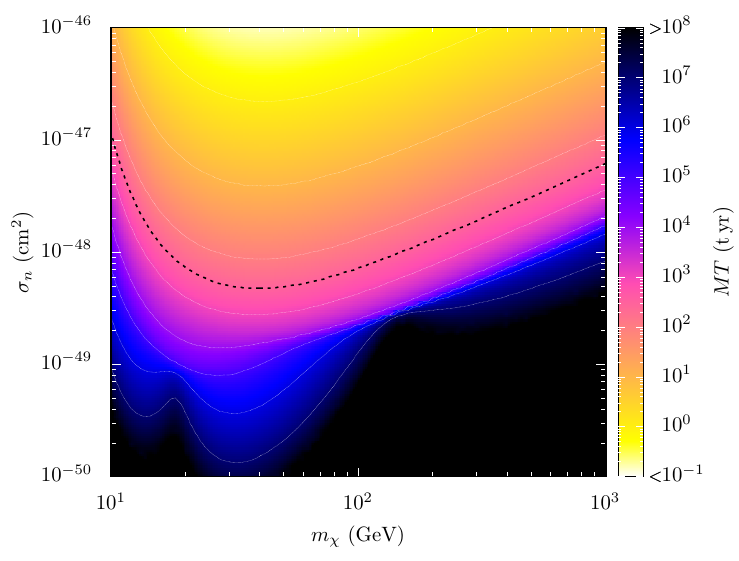}
  \includegraphics[width=.47\textwidth]{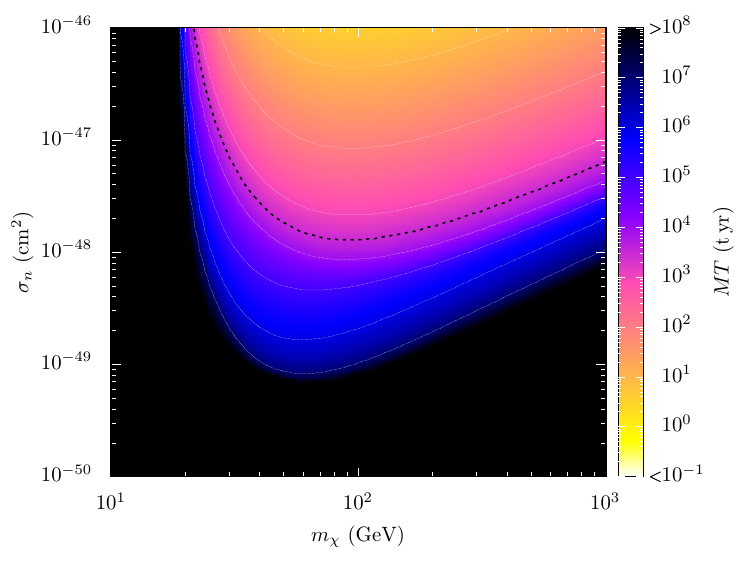}
  \caption{Dark matter discovery limits in xenon~(left) and argon~(right)
detectors for the baseline detector parameters summarized in Tables~\ref{tab:flux} and \ref{tab:detector}. The black dashed lines indicate the sensitivities for a DARWIN-like experiment with exposure $MT = 3\times 10^2 \text{t\,yr}$~(left) and for an ARGO-like experiment with $MT = 3\times 10^3\,\text{t\,yr}$~(right).}
  \label{fig:contours}
\end{figure}

 We turn now to applying the methods discussed above to estimate the effect of neutrino backgrounds on the dark matter discovery sensitivities of future large-scale liquid argon and experiments. In particular, we investigate the impact of atmospheric flux uncertainties, electron-recoil rejection factors, and recoil energy sensitivity. We also study the potential gain in dark matter sensitivity that might be obtained by combining data from argon and xenon experiments.  

 For the baseline flux and detector parameters used in our analysis listed in Tables~\ref{tab:flux} and \ref{tab:detector}, the SI dark matter discovery limits we find are summarized in Fig.~\ref{fig:contours} for xenon~(left) and argon~(right). Our analysis is general, but we highlight in particular the sensitivities of a DARWIN-like xenon experiment with a total exposure of $MT = 2\times 10^2\,\mathrm{t\,yr}$ and an ARGO-like argon experiment with exposure of $MT = 3\times 10^{3}\,\mathrm{t\,yr}$~\cite{Billard:2021uyg}. These are indicated by dashed black lines in the respective panels of Fig.~\ref{fig:contours}. At very large exposures, the sensitivity in xenon shows two distinctive features near $m_\chi \sim 20\,\gev$ and $m_\chi \sim 150\,\gev$. For these masses, the energy spectra from dark matter scattering match very closely with the neutrino-nucleus spectra primarily from the DSNB ($m_\chi \sim 20\,\gev$) and from atmospheric sources ($m_\chi \sim 150\,\gev$) (shown in Fig.~\ref{fig:spectra}).  In the rest of the section, we investigate the effects of varying away from these baselines.

\subsection{Impact of Atmospheric Fluxes and Uncertainties}

Atmospheric neutrinos are the dominant source of neutrino-nucleus scattering background for weak-scale dark matter discovery. The nuclear recoil energy spectrum they induce can be very similar to dark matter scattering, particularly for certain specific dark matter masses. The ability of future dark matter detectors to distinguish dark matter from atmospheric neutrinos through spectral shape information is therefore very sensitive to how well the atmospheric neutrino flux is known. Uncertainty on the net flux in the relevant energy range, summed over all flavours of neutrinos and antineutrinos, is estimated in Refs.~\cite{Battistoni:2005pd,Honda:2006qj,Honda:2015fha} to be approximately $20\%$.

\begin{figure}[ttt]
  \centering
  \includegraphics[width=.47\textwidth]{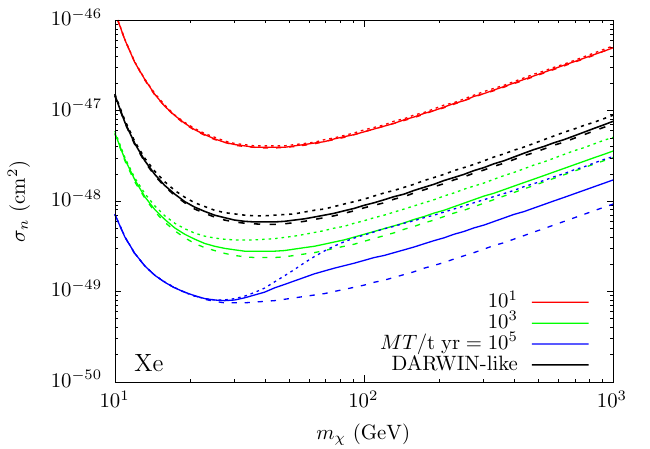}
  \includegraphics[width=.47\textwidth]{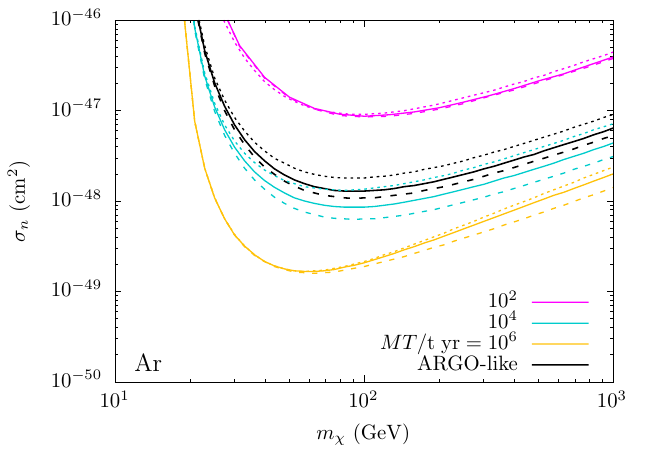}\\
  \includegraphics[width=.47\textwidth]{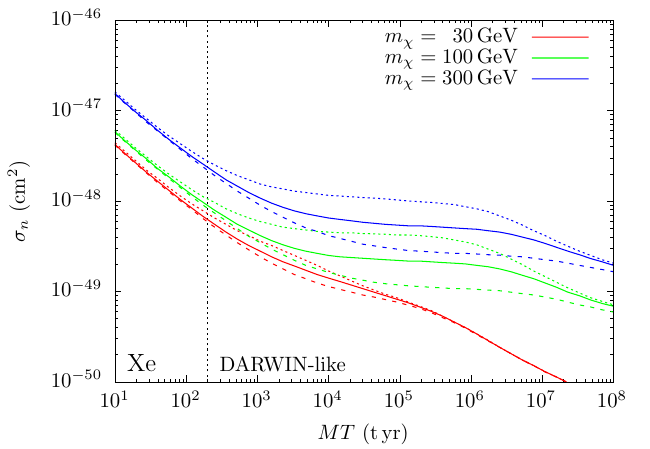}
  \includegraphics[width=.47\textwidth]{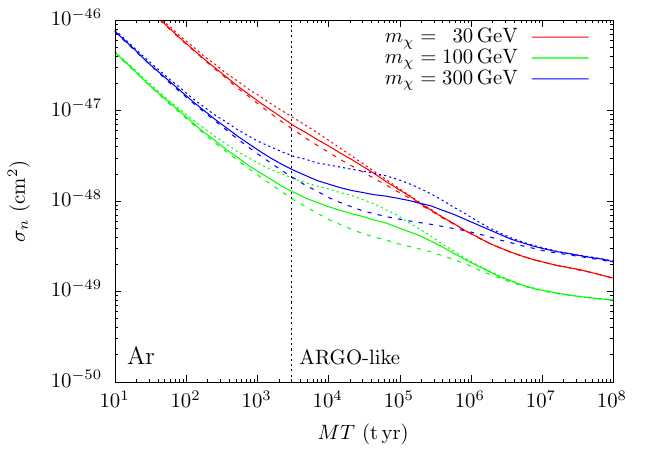}
  \caption{Effect of atmospheric neutrino flux uncertainties on the dark matter discovery limits of xenon~(left) and argon~(right) detectors as a function of dark matter mass~(top) for the listed fixed values of exposure $MT/(\text{t\,yr})$ and as function of exposure for $m_{\chi} = 30,\,100,\,300\,\gev$~(bottom). The solid lines indicate the sensitivity with the current uncertainty estimate on the total flux of $20\%$, while the dotted~(dashed) lines show the sensitivity for a flux uncertainty of $40\%$~($10\%$).}
  \label{fig:atmspec}
\end{figure}

To demonstrate the impact of this uncertainty, we show in the upper panels of Fig.~\ref{fig:atmspec} the dark matter discovery limit as a function of mass $m_{\chi}$ for various fixed values of exposure $M\,T$ in xenon~(left) and argon~(right). Similarly, in the lower panels we show the discovery limits in xenon~(left) and argon~(right) as a function of exposure $M\,T$ for masses $m_{\chi} = 30,\,100,\,300\,\gev$ in Fig.~\ref{fig:atmspec}. In both figures, the solid lines correspond to a fractional total atmospheric flux uncertainty of $20\%$, while the dotted lines denote $40\%$ and the dashed lines $10\%$. The $20\%$ line represents the current best uncertainty estimate, the $10\%$ line demonstrates the effect of improved atmospheric flux determinations on dark matter sensitivity, while the $40\%$ lines illustrate how much worse the dark matter sensitivity would get if the atmospheric background was not as well determined. 
 
The importance of the uncertainty in the atmospheric neutrino flux has a significant impact on the dark matter discovery reach in both xenon and argon detectors. The effect is important both for smaller dark matter cross-sections where the total event rate is dominated by atmospheric neutrinos, as well as for larger dark matter masses where the recoil energy spectrum from dark matter becomes similar to that from atmospheric neutrinos. Improved measurements of the atmospheric neutrino flux at energies below $E_{\nu} \sim 100\,\mev$ could therefore increase the reach of future large-scale dark matter searches. Projections for DUNE \cite{Kelly:2019itm} suggest that a fractional uncertainty on the order of $10\%$ may be achievable in this energy range. However, for DARWIN- or ARGO-scale detectors, the impact of this improvement appears to be fairly modest. An additional study of the effect of uncertainties in the shape of the atmospheric neutrino energy spectrum is presented in Appendix~\ref{sec:appb} with a similar conclusion.

Plotting the dark matter discovery reach as a function of $MT$, as shown in Fig.~\ref{fig:atmspec}, also illustrates the structure of the neutrino floor phenomenon. As the neutrino background becomes so large that its uncertainty grows larger than the dark matter signal, increasing the exposure $MT$ further yields almost no improvement. This effect is more pronounced in xenon than in argon since the signal and background energy spectra tend to be more similar for xenon. As the exposure is increased even further, to values beyond currently foreseeable capabilities, we find that the background shape is determined so well from data that the dark matter sensitivity starts increasing again, as $\sqrt{MT}$. In effect, there are enough neutrino events in this regime for the background to be determined directly from the data rather than the input uncertainty. This ''softness'' of the neutrino floor was observed in Ref.~\cite{Billard:2013qya,Ruppin:2014bra} and studied in Ref.~\cite{OHare:2020lva}. We will investigate it further below in relation to other sources of uncertainties. 

The magnitude and energy spectrum of the atmospheric flux also varies with time and location. The time variation comes mainly from the 11-year solar cycle since an increased solar activity tends to deflect more cosmic rays from the solar neighbourhood. Our estimates are based on the calculation of Ref.~\cite{Battistoni:2005pd} which uses the solar-cycle average for the incident cosmic ray flux spectrum. A multi-year experiment could potentially characterize (and reduce the impact of) the atmospheric neutrino background more effectively by taking this variation into account (and see Ref.~\cite{OHare:2020lva} for a related discussion). Atmospheric neutrino fluxes also vary by location (primarily latitude) on account of the geomagnetic rigidity cutoff on lower-energy cosmic rays imposed by the Earth's magnetic field. Our atmospheric flux estimates based on Ref.~\cite{Battistoni:2005pd} are for a detector located at LNGS at the mean of the solar cycle. 

\begin{figure}[ttt]
  \centering
  \includegraphics[width=.47\textwidth]{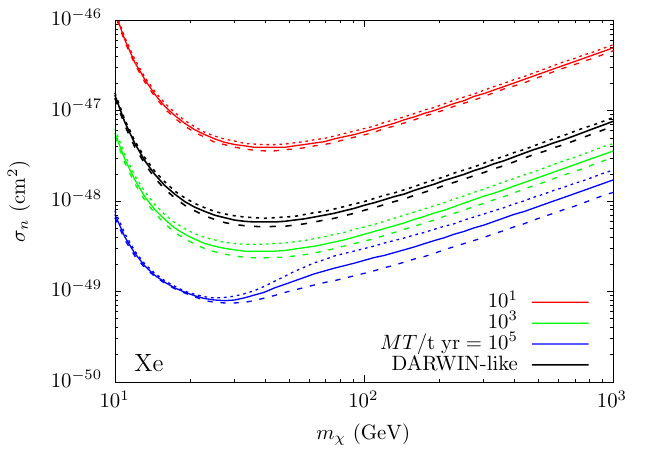}
  \includegraphics[width=.47\textwidth]{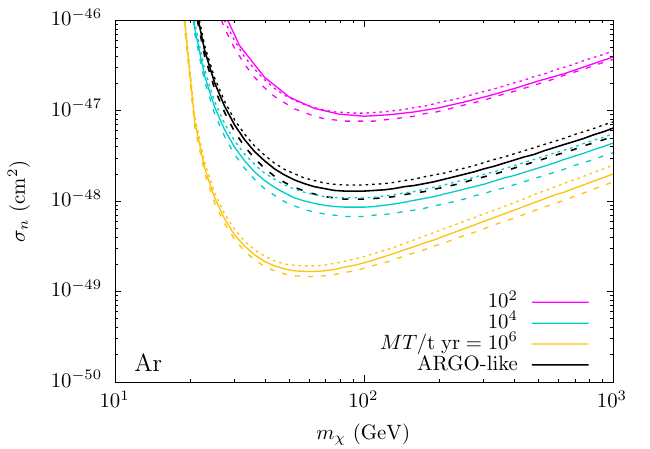}
  \caption{Estimated effect of different detector locations on the dark matter discovery sensitivity in xenon~(left) and argon~(right) detectors based on varying the overall magnitude of the atmospheric flux for the listed values of exposure $MT/(\text{t\,yr})$. The solid lines show the dark matter sensitivity using the flux calculated for LNGS, the dotted lines show the sensitivity for a relative 30\% increase in atmospheric flux as an estimate of the flux at SNOLAB, and the dashed lines show the sensitivity for a relative 30\% decrease in atmospheric flux as an estimate for Super-Kamiokande.
  }
  \label{fig:atmsite}
\end{figure}

To estimate the impact of detector location on the sensitivity to dark matter, we show the effect of a 30\% increase~(decrease) in the overall magnitude of the atmospheric neutrino flux on the dark matter discovery power with the dotted~(dashed) lines in Fig.~\ref{fig:atmsite} for xenon~(left) and argon~(right). The solid lines show the baseline atmospheric flux at LNGS from Ref.~\cite{Battistoni:2005pd}. Based on the neutrino flux predictions of Ref.~\cite{Honda:2006qj} which are given for LNGS ($42.45^{\circ}$N latitude), Super-Kamiokande ($36.43^{\circ}$N latitude), and SNOLAB ($46.47^{\circ}$N latitude) for neutrino energies $E_{\nu} \geq 100\,\mev$, and those of Ref.~\cite{Battistoni:2005pd} given for LNGS and Super-Kamiokande down to $E_{\nu} \geq 10\,\mev$, this simple rescaling of the LNGS flux gives a very approximate estimate of the fluxes at SNOLAB~(30\% larger) and Super-Kamiokande~(30\% smaller). The impact is found to be moderate for realistic exposures. However, these results also show that future large-scale dark matter direct searches would benefit from updated calculations of the atmospheric neutrino flux down to $E_{\nu} \sim 10\,\mev$ that are specific to the location of the detector and adjusted in time to account for solar activity and atmospheric variations.

\subsection{Impact of Electron Recoil Rejection}

Neutrino-electron scattering is a background to dark matter-nucleus scattering due to the finite ability of detectors to distinguish between nuclear and electron recoils. The energy transfer from neutrinos to electrons is more efficient than for nuclear recoils, and thus lower energy but high-flux $pp$ and $^7$Be solar neutrinos can produce electron recoil energies in the dark matter search regions of interest.  Single-phase argon detectors have demonstrated excellent electron recoil rejection better than $\varepsilon_e \lesssim 10^{-8}$ through pulse shape discrimination~\cite{Ajaj:2019imk,Aalseth:2017fik,Adhikari:2021wrm},  and we find that this is sufficient to remove neutrino-electron recoils as relevant background for realistic future detectors. However, electron recoil rejection is a greater challenge in xenon-based experiments. The stated nominal goal for the DARWIN detector of $\varepsilon_e = 2\times 10^{-4}$ will reduce but not eliminate this background.

\begin{figure}[ttt]
  \centering
  \includegraphics[width=.47\textwidth]{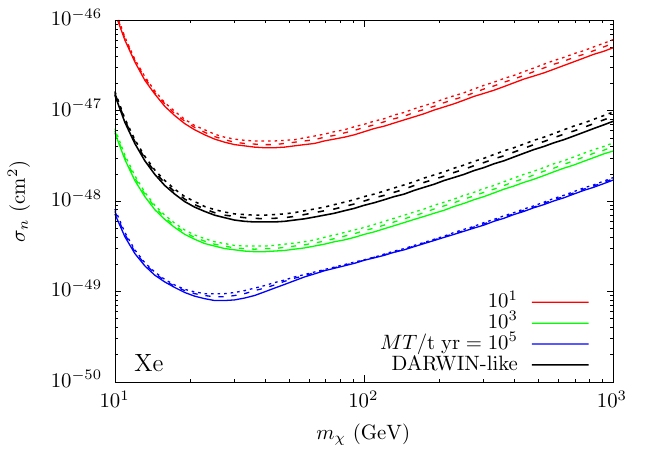}
  \includegraphics[width=.47\textwidth]{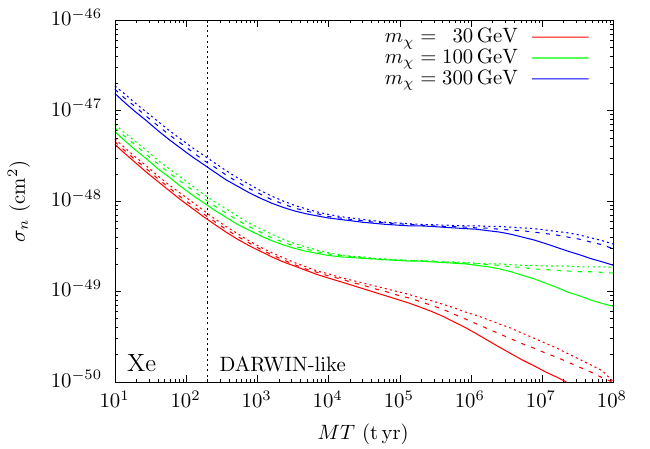}
  \caption{Impact of the electron rejection power $\varepsilon_e$ on the dark matter discovery sensitivity in xenon as a function of dark matter mass~(left) for the listed values of the exposure $MT/(\text{t\,yr})$, and of exposure~(right) for $m_{\chi} = 30,\,100,\,300\,\gev$. The solid lines correspond to the DARWIN goal of $\varepsilon_e = 2\times 10^{-4}$ while the dashed~(dotted) lines show $\varepsilon_e = 1\times 10^{-3}$~($2\times 10^{-3}$).}
  \label{fig:erfxe0}
\end{figure}

We illustrate the impact of electron recoil rejection in xenon in Fig.~\ref{fig:erfxe0}, where we show the dark matter discovery sensitivity as a function of the dark matter mass~(left) and the exposure~(right). The solid lines correspond to the DARWIN goal of $\varepsilon_e = 2\times 10^{-4}$, while the dashed~(dotted) lines show $\varepsilon_e = 1\times 10^{-3}$~($2\times 10^{-3}$). Less efficient electron rejection (larger $\varepsilon_e$) reduces the discovery sensitivity, but we find that the effect is relatively mild for realistic exposures and $\varepsilon_e \lesssim 2\times10^{-3}$. For planned future xenon experiments, this suggests that increasing the electron rejection power might not improve the sensitivity to dark matter if it also leads to a significant decrease in nuclear recoil acceptance~\cite{Schumann:2015cpa}. In contrast, at much larger exposures, a reduced electron rejection power can lead to significant degradation of the dark matter reach and solidification of the neutrino floor.

So far, we have treated the electron rejection factor $\varepsilon_e$ as constant as a function of recoil energy with no uncertainty. In practice, however, this quantity will have an energy-dependent uncertainty both from the characterization of a given detector~\cite{Akerib:2020lkv,Aprile:2017xxh}, as well as the challenge in relating the energies of electron recoils to nuclear recoils~\cite{Szydagis:2021hfh}. To investigate the potential effects of not knowing $\varepsilon_e$ exactly, we model this uncertainty in two simple ways. First, we allow for a $20\%$ energy-independent uncertainty relative to the central value of $\varepsilon_e$. And second, we allow $\varepsilon_e$ to vary independently in each energy bin by $10\%$ relative to the central value of $\varepsilon_e$ to approximate the effect of uncertainty in the energy dependence of the electron rejection. For both uncertainty models, we treat the relevant variations as nuisance parameters in our profile likelihood analysis following the same approach as for the neutrino flux uncertainties. 
 
\begin{figure}
  \centering
  \includegraphics[width=.47\textwidth]{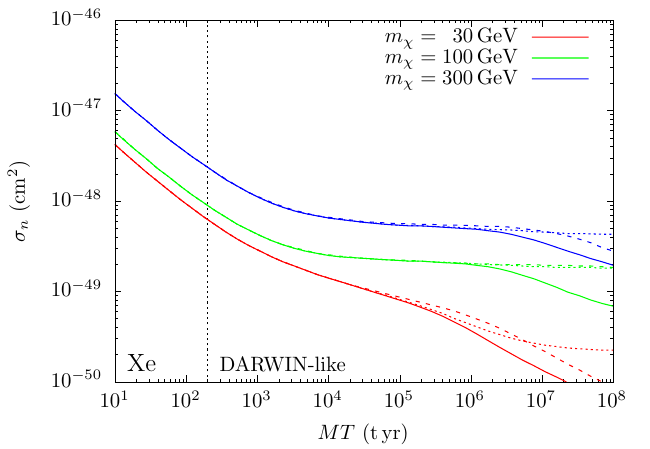}
  \includegraphics[width=.47\textwidth]{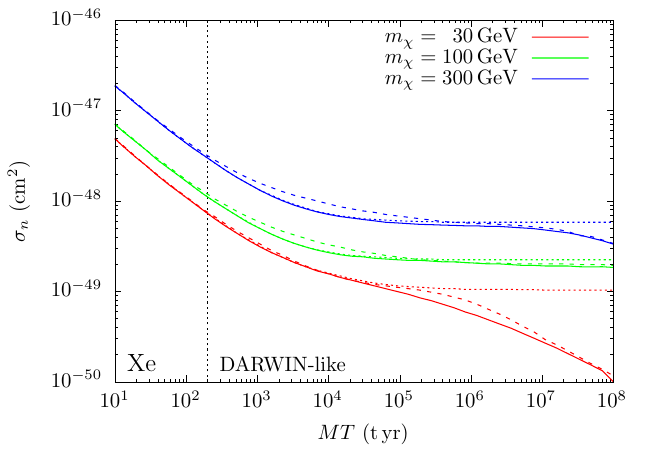}
  \caption{Effect of uncertainties in electron recoil rejection power on the dark matter discovery sensitivity in xenon as a function of exposure $MT$ for dark matter masses $m_{\chi}=30,\,100,\,300\,\gev$. The left~(right) panel has a baseline electron rejection factor of $\varepsilon_e=2\times 10^{-4}$~($2\times 10^{-3}$), with the solid lines showing the sensitivity with no uncertainty in $\varepsilon_e$, the dashed lines giving the sensitivity for an overall $20\%$ uncertainty on $\varepsilon_e$, and the dotted lines showing the sensitivity with a $10\%$ bin-by-bin uncertainty on $\varepsilon_e$.
  }
  \label{fig:erfxe3}
\end{figure}

Results for these two electron recoil rejection uncertainty models are given in Fig.~\ref{fig:erfxe3}, where we show the dark matter discovery reach for $m_{\chi} = 30,\,100,\,300\,\gev$ as a function of exposure $M\,T$. In the left panel, we use a baseline electron rejection of $\varepsilon_e = 2\times 10^{-4}$, while in the right panel we have a baseline of $\varepsilon_e = 2\times 10^{-3}$. The solid lines in both panels reproduce the previous results with no uncertainty in $\varepsilon_e$ (shown in Fig.~\ref{fig:erfxe0}), while the dashed lines show the $20\%$ overall uncertainty model and the dotted lines indicate the 10\% independent energy bin model. Not surprisingly, uncertainties in the electron recoil rejection degrade the sensitivity to dark matter. More notably, however, the energy-dependent uncertainty of the second model leads to a much more rigid neutrino floor, even for dark matter masses where the dark matter and neutrino-electron recoil spectra are significantly different. When the shape of electron recoil background is known, it can be determined very well with enough data from the highest energies measured, where it dominates over nuclear scattering (as shown in Fig.~\ref{fig:spectra}). It can then be extrapolated down to the lower energies where the dark matter signal is expected to lie. However, such an extrapolation is impeded by energy-dependent uncertainties since the electron scattering background at lower energies is no longer completely fixed completely by its value at higher energies. 
 
In future experiments, it would be helpful for the collaborations to publish data-driven estimates for the uncertainties in electron recoil rejection as a function of the reconstructed recoil energy $E_R$. This result illustrates a more general point that recoil-energy-dependent uncertainties reduce the ability of the profile likelihood to learn the background distributions.

\subsection{Dependence on the Recoil Energy ROI}
Our analysis has assumed dark matter nuclear recoil energy search regions motivated by projections for future large-scale noble element detectors. For xenon, this range is $E_R \in [5,35]\,\kev$ while for argon it is $E_R\in [55,100]\,\kev$. In both cases, these regions of interest~(ROIs) are chosen to maximize dark matter acceptance while accounting for expected detector sensitivities and mitigating potential (non-neutrino) backgrounds. Even so, it also potentially useful to understand the effect of expanding these search regions. 

\begin{figure}[ttt]
  \centering
  \includegraphics[width=.47\textwidth]{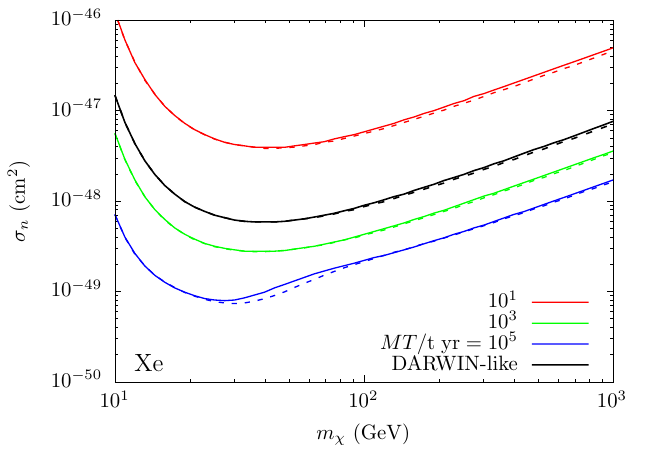}
  \includegraphics[width=.47\textwidth]{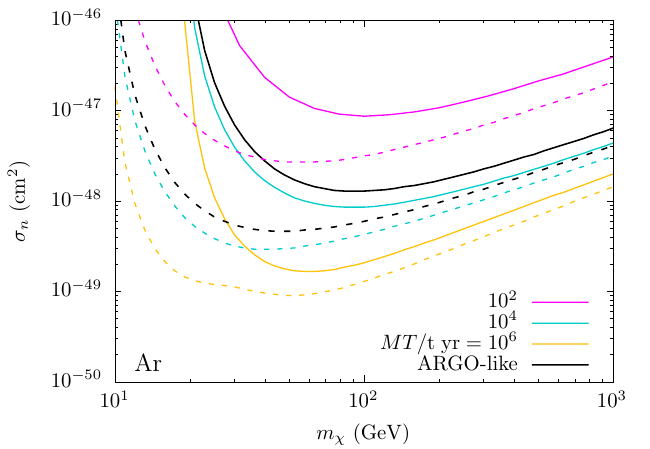}
  \caption{Impact of expanding the recoil energy ROI on the dark matter discovery sensitivity in xenon~(left) and argon~(right) as a function of mass for a set of fixed exposures $MT/(\text{t\,yr})$. The solid lines show the sensitivity for our fiducial ROIs, $E_R = [5,35]\,\kev$ for xenon and $E_R\in [55,100]\,\kev$ for argon. The dashed lines indicate the sensitivity when these are expanded to $E_R\in [5,100]\,\kev$ for xenon and $E_R\in [20,100]\,\kev$ in argon.}
  \label{fig:roispec}
\end{figure}

 In Fig.~\ref{fig:roispec} we show the effect on the dark matter discovery sensitivity of increasing the ROIs in xenon from $E_R\in [5,35]\,\kev$ to $E_R\in [5,100]\,\kev$~(left), and in argon from $E_R\in [55,100]\kev$ to $E_R\in [20,100]\,\kev$. For xenon, the lower bound is limited by the flux of solar $^8B$ neutrinos. However, the upper bound could be raised with improvements in both the electron recoil rejection and nuclear recoil acceptance~\cite{Schumann:2015cpa}. Even if this can be achieved with no significant increase in other backgrounds, it does not improve the dark matter discovery reach very much. At these larger recoil energies, both the dark matter and neutrino-nucleus scattering rates are strongly reduced by the small values of the nuclear form factor in this range. In contrast, expanding the ROI in argon down to $E_R$ as low as $20\,\kev$ produces a large increase in the sensitivity to dark matter. This is to be expected for lower dark matter masses below $m_{\chi}\lesssim 100\,\gev$ where the maximum recoil energy is not very large, but the increase also extends to much larger masses. At such larger masses, a significant portion of the nuclear recoil spectrum lies below $E_R = 55\,\kev$, so expanding the ROI at the low end also yields more dark matter events. To achieve a lower energy threshold in a (single-phase) argon detector would require a strong increase in the overall photo-collection efficiency, pushing to the statistical limitations of the intrinsic scintillating properties of argon~\cite{Adhikari:2020zyy,Boulay:2004dk}.

\subsection{Combining Data from Argon and Xenon Experiments}

Dark matter and neutrino recoil energy spectra are significantly different in argon relative to xenon, as shown in Fig.~\ref{fig:spectra}. Furthermore, the dominant uncertainties in the neutrino backgrounds are strongly correlated between different experiments. These features suggest that combining data from future argon and xenon experiments might enhance the sensitivity to dark matter beyond just increasing the total effective exposure. We investigate this possibility here.  A similar study using other detector materials was done in Ref.~\cite{Ruppin:2014bra}. 

The natural generalization of the discovery test statistic discussed in Sec.~\ref{sec:detectors} is to update $q_0$ from Eq.~\eqref{eq:qzero} with the combined likelihood function
\beq \label{eq:combine}
\mathcal{L}(\xi,\vec{\theta}) = 
\prod_{i=1}^{M_b}\mathcal{P}(M^i;\langle M^i\rangle)
\times
\prod_{j=1}^{N_b}\mathcal{P}(N^j;\langle N^j\rangle)
\times
\mathcal{L}_b(\vec{\theta}) \ ,
\eeq
where $M^i$ is the number of events in the $i$-th bin of the argon experiment and $N^j$ is the number of events in the $j$-th bin of the xenon experiment. The key feature of this combined likelihood is that the same neutrino background parameters apply to both sets of data. This is certainly true for solar and DSNB neutrinos, and also for atmospheric neutrinos provided the two detectors are at the same location. We also expect it to be a reasonable approximation for the atmospheric fluxes for detectors at different locations since their dominant uncertainties, the underlying interaction model and the primary cosmic ray flux spectrum, are strongly correlated~\cite{Battistoni:2005pd,Honda:2006qj,Honda:2011nf,Honda:2015fha}.

\begin{figure}[ttt]
  \centering
  \includegraphics[width=.47\textwidth]{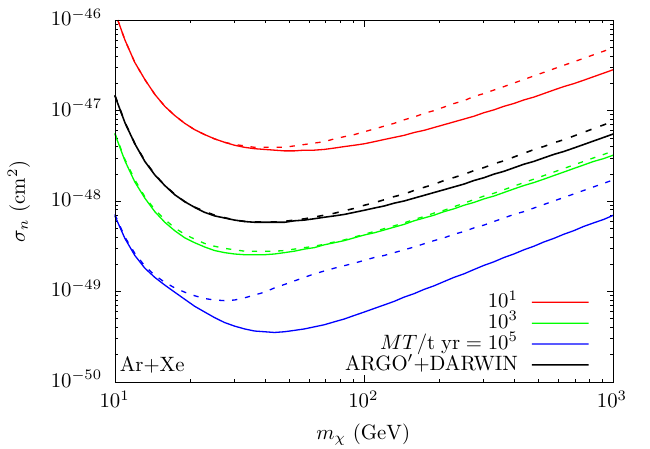}
  \includegraphics[width=.47\textwidth]{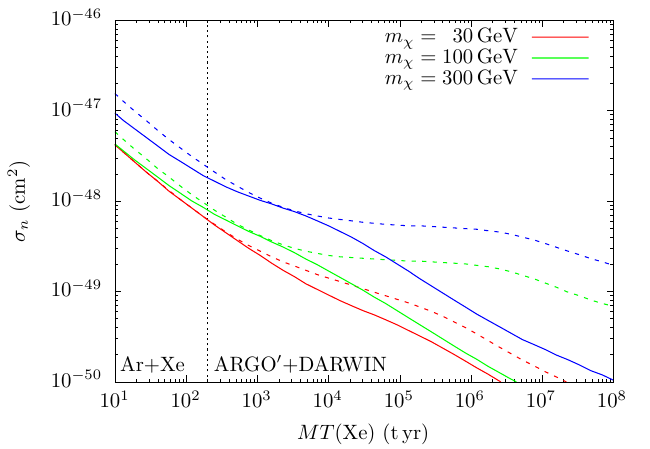}
  \caption{Dark matter discovery sensitivities obtained by combining results from a xenon experiment with exposure $MT$ with an argon experiment with exposure $MT_{\text{Ar}}= 10\,MT$ as a function of mass~(left) for the listed fixed values of xenon exposure $MT/(\text{t\,yr})$, and of xenon exposure~(right) for $m_{\chi}=30,\,100,\,300\,\gev$. The solid lines show the combined sensitivity while the dashed lines indicate the sensitivity of a xenon experiment alone.}
  \label{fig:comb}
\end{figure}

 The estimated result for the dark matter discovery reach of combining data from independent xenon and argon experiments is presented in Fig.~\ref{fig:comb} as a function of dark matter mass~(left) and total xenon exposure $MT$~(right). The solid lines show the combined sensitivity, while the dashed lines show the result for xenon alone. In both panels, the total argon detector exposure is taken to be ten times the xenon exposure, $MT_{\text{Ar}}= 10\,MT$. This is in line with the expected relative sizes of future detectors. All other detector parameters such as ROI and electron recoil rejection are set to the reference values used previously. 

The enhancement of the dark matter discovery reach from combining argon and xenon data is striking at very large exposures. This combined sensitivity is much greater than either experiment individually. With only a single detector material, to a significant extent, it is possible to make up for the lack of dark matter in the background-only hypothesis by adjusting the neutrino flux parameters. When argon and xenon data are combined, this compensation effect is much less effective due to the different recoil energy spectra in these two materials owing to argon being much lighter than xenon. For exposures in line with proposed next-generation experiments, the increase in sensitivity from combining argon and xenon data is more modest. The increased combined sensitivity is also likely to be moderated by other background sources and uncertainties related to the individual detectors. Even so, the power of combining data from lighter and heavier detector materials offers a potential brute force approach to testing dark matter beyond the (not so rigid) neutrino floor.

\section{Conclusions\label{sec:conc}}

In this paper, we investigated the nature of neutrino backgrounds for future large-scale liquid argon and xenon dark matter experiments. We reviewed the calculation of scattering rates of dark matter and neutrino scattering on nuclei and electrons in the target material in Section~\ref{sec:scatt}. Next, in Section~\ref{sec:detectors} we connected these scattering rates to the properties of expected future detectors and the statistical methods used to study them. We then applied these methods in Section~\ref{sec:nufloor} to investigate the effect of neutrino scattering on the dark matter discovery sensitivity of such detectors.

 Our results confirm previous studies of neutrino backgrounds to dark matter discovery in xenon detectors based on a profile likelihood analysis~\cite{Billard:2013qya,Ruppin:2014bra,Gelmini:2018ogy,OHare:2020lva}. As these backgrounds approach the dark matter signal being looked for, the scaling of the discovery sensitivity with exposure $M\,T$ goes from linear to square root. When the systematic uncertainties in the background rates approach the size of the signal being looked for, the scaling with exposure slows even further. This is commonly referred to as the \emph{neutrino floor}. However, if the recoil energy spectra of these backgrounds are known precisely, with enough additional data, a dark matter signal can be distinguished from neutrinos through spectral information and scaling with the square root of exposure resumes~\cite{Billard:2013qya,Ruppin:2014bra,OHare:2020lva}.

 The most important neutrino background in searches for WIMP(-like) dark matter with mass $m_{\chi}\gtrsim 10\,\gev$ comes from atmospheric neutrinos. The search reach of future detectors is therefore sensitive to how well the total atmospheric flux can be determined, particularly in the neutrino energy range $E_{\nu} \in [30,300]\,\mev$. We find that direct measurements of this flux could be extremely valuable to future dark matter searches. Updated calculations of the atmospheric neutrino flux as functions of the time and location of dark matter searches would also be helpful.

 A further important source of backgrounds is sub-MeV energy solar neutrinos scattering off atomic electrons in dark matter detectors. Pulse shape information can be used in argon to distinguish this source from dark matter scattering on nuclei~\cite{Ajaj:2019imk}. Distinguishing electron recoils is a greater challenge in xenon, and future detectors are expected to have electron rejection factors in the range $\varepsilon_e \simeq 2\times 10^{-4}$--$2\times 10^{-3}$~\cite{Akerib:2018lyp,Zhang:2018xdp,Aprile:2020vtw,Aalbers:2016jon}. Our results suggest that this level of rejection is largely sufficient to handle neutrino-electron scattering backgrounds for the current and next generation of xenon detectors with total exposures up to $MT \sim 200\,\text{t\,yr}$. This is possible, in part, because the recoil energy spectrum from neutrino-electron scattering is significantly different from that of nuclear recoils, allowing for this background to be determined from data. However, we also note that if the shape of the background is uncertain, its determination from data becomes much more difficult.

Detectors can also be characterized by the recoil energy range over which they are sensitive to dark matter. For xenon, we assumed a region of interest for nuclear recoils of $E_R\in [5,35]\,\kev$. This range does a good job of covering dark matter masses above about $10\,\gev$; extending this region to lower energies introduces a large neutrino background from $^8$B and hep neutrinos, while the nuclear scattering rate at higher energies is suppressed by the nuclear form factor. For argon experiments, we assumed a search region of $E_R\in [55,100]\,\kev$. The lower threshold here comes from the requirement of obtaining enough photons to reject electron events efficiently through pulse shape discrimination. Our results show that lowering this threshold energy would lead to more dark matter scattering events and could increase significantly the dark matter sensitivity of argon experiments, even for very heavy dark matter.

 The planned future program of large-scale dark matter detectors is expected to include experiments based on both argon and xenon. These two detector materials have different nuclear recoil energy spectra, both for dark matter and neutrinos. Based on this feature, we find that combining data from argon and xenon experiments can yield a much greater sensitivity than either one individually. (See also Refs.~\cite{Ruppin:2014bra} for a similar analysis for other detector materials.) While a dark matter signal in a detector of one material can be compensated for to a large extent by the uncertainties in the neutrino backgrounds (particularly in xenon), this is much more difficult to achieve simultaneously in argon and xenon experiments together. The enhancement in sensitivity is modest for the next generation of xenon and argon detectors with exposures up to $M\,T \sim 2\times 10^2\,\text{t\,yr}$ and $3\times 10^3\,\text{t\,yr}$, respectively. However, the increase in sensitivity at higher exposures is striking, and combining data from argon and xenon experiments could provide a brute force way to go beyond the neutrino floor.

 Planned large-scale dark matter searches in argon and xenon, such as ARGO and DARWIN, will have to contend with neutrino scattering as an important background. Our results suggest that the impact of these backgrounds on the dark matter discovery reach is sensitive to the atmospheric neutrino flux and the uncertainties in it while depending less strongly on reasonable variations (and uncertainties) in the electron rejection power and the search regions of interest. If systematic uncertainties can be controlled, our results also indicate that it may be possible to reach beyond the so-called neutrino floor by combining data from even larger argon and xenon experiments.

\begin{acknowledgments}
We thank Austin de Ste Croix, C.~Eric Dahl, Gilly Elor, Rafael Lang, David McKeen, Scott Oser, Nirmal Raj, Fabrice Reti\`ere, and Alan Robinson for helpful discussions and comments provided during this project. The work of Pietro Giampa is supported by the by the Arthur B. McDonald  Canadian  Astroparticle  Physics  Research  Institute, while David E. Morrissey is supported by the Natural Sciences and Engineering Research Council of Canada~(NSERC). SNOLAB operations are supported by the Canada Foundation for Innovation~(CFI) and the Province of Ontario Ministry of Research and Innovation. TRIUMF receives federal funding via a contribution agreement with the National Research Council~(NRC) of Canada. 
\end{acknowledgments}

\appendix

\section{Evaluating the Halo Integral \label{sec:appa}}

As demonstrated in Eq.~\eqref{eq:dmrate2}, all the dependence of  the dark matter scattering rate on the astrophysical dark matter distribution can be collected into the local dark matter energy density $\rho_{\chi}$ and the so-called halo integral~\cite{Lewin:1995rx,Jungman:1995df},
\beq
\eta(v_{min}) \ = \ \int_{v_{min}}\!\!d^3v\;\frac{f_{lab}(\vec{v})}{v} \ ,
\eeq
 where $\vec{v}$ is the dark matter velocity in the lab frame, $f_{lab}$ is the lab-frame velocity distribution, and $v_{min}(E_R) = \sqrt{m_NE_R/2\mu_N^2}$ for a given recoil energy $E_R$. The lab-frame velocity distribution is related to the local distribution in the galactic halo frame $f(\vec{v}^{\,\prime})$ by
\beq
f_{lab}(\vec{v}) = f(\vec{v}^{\,\prime}) = f(\vec{v}+\vec{v}_{E}) \ ,
\eeq
 where $\vec{v}^{\,\prime}$ is the dark matter velocity in the halo frame, and $\vec{v}_E$ is the net velocity of the Earth relative to the dark matter halo.

 To evaluate the halo integral, we follow the recent recommendations of Ref.~\cite{Baxter:2021pqo} and set $\rho_{\chi} = 0.3\,\gev\,\mathrm{cm}^{-3}$ for the local dark matter density together with the Standard Halo Model~(SHM) velocity distribution, 
\beq
f(v) = \mathcal{N}\,\lrf{1}{\pi\,v_0^2}^{3/2}e^{-v^2/v_0^2}\;\Theta(\vesc-v) \ .
\label{eq:fvshm}
\eeq
 The factor of $\mathcal{N}$ is a normalization factor given by
\beq
\mathcal{N}^{-1} = \erf(z_{esc}) - \frac{2}{\sqrt{\pi}\,}\,z_{esc}\,e^{-z_{esc}^2} \ ,
\nnmb 
\eeq
with $z_{esc} = \vesc/v_0$, and
\beq
\erf(x) = \frac{2}{\sqrt{\pi}\,}\int_0^{x}\!dt\,e^{-t^2} \ ,
\eeq
 is the usual error function. In our analysis we use the recommended SHM parameters from Ref.~\cite{Baxter:2021pqo}: $v_0 = 238\,\text{km/s}$, $v_E = 254\,\text{km/s}$, and $v_{esc} = 544\,\text{km/s}$.

 A convenient feature of the SHM is that it allows for an analytic expression for the halo integral defined in Eq.~\eqref{eq:haloint}. For $\vesc > v_0, v_E$, the result is
\beq
\eta(\vmin) = \frac{\mathcal{N}}{v_0}\left\{
\begin{array}{lcl}
0&;& \vmin > (\vesc\! +\! v_E)\\
\frac{\erf(z_{min}+z_E) - \erf(z_{min} - z_E)}{2z_E}
- \frac{2}{\sqrt{\pi}}e^{-z_{esc}^2}
&;& \vmin < (\vesc\!-\! v_E)
\hspace{-3cm}
\\
\frac{\erf(z_{esc}) - \erf(z_{min} - z_E)}{2z_E}
&;& (\vesc\!-\!v_E) < \vmin < (\vesc\!+\!v_E)\\
~~~~~- \frac{1}{\sqrt{\pi}}\lrf{z_{esc}+z_E-z_{min}}{z_E}\,e^{-z_{esc}^2}&&
\end{array}\right.
\eeq
where $z_i = v_i/v_0$.

\section{Shape Uncertainties in the Atmospheric Flux Spectrum\label{sec:appb}}

In the analysis above, we showed that atmospheric neutrinos are typically the most important contribution to the neutrino background to dark matter detection. This analysis assumed an uncertainty in the normalization of the atmospheric neutrino energy spectrum with the spectral shape held fixed based on the calculation of Ref.~\cite{Battistoni:2005pd}. However, the energy dependence of the atmospheric flux also has an uncertainty. We investigate the effect of such an uncertainty on dark matter discovery in this appendix.

 As a first step, it is helpful to connect specific atmospheric neutrino energies to the nuclear scattering recoil energies they induce. Collecting neutrino energies according to
\beq
E_k  =  10\,\mev\lrf{10^4\,\mev}{10\,\mev}^{k/N_\nu} \ ,
\label{eq:enubins}
\eeq
 we define $k=1,2,\ldots\,N_\nu = 20$ neutrino energy bins with range $E_{\nu}\in [E_{k-1},E_k]$. With these in hand, consider now the quantities
\beq
\mu^{ik} = n_N\,\int_{E_{i-1}}^{E_i}\!dE_R\,\int_{E_{k-1}}^{E_k}\!dE_\nu\;
\phi_{atm}(E_{\nu})\,\frac{d\sigma_{\nu N}}{dE_R} \ .
\label{eq:muik}
\eeq
 For each recoil energy bin $i$, $\mu^{ik}$ is the contribution to that recoil bin per unit exposure from atmospheric neutrinos with energies in the $k$-th neutrino bin. Multiplying by exposure $MT$ would then produce the expected number of events from these energy ranges.

\begin{figure}[ttt]
  \centering
  \includegraphics[width=.47\textwidth]{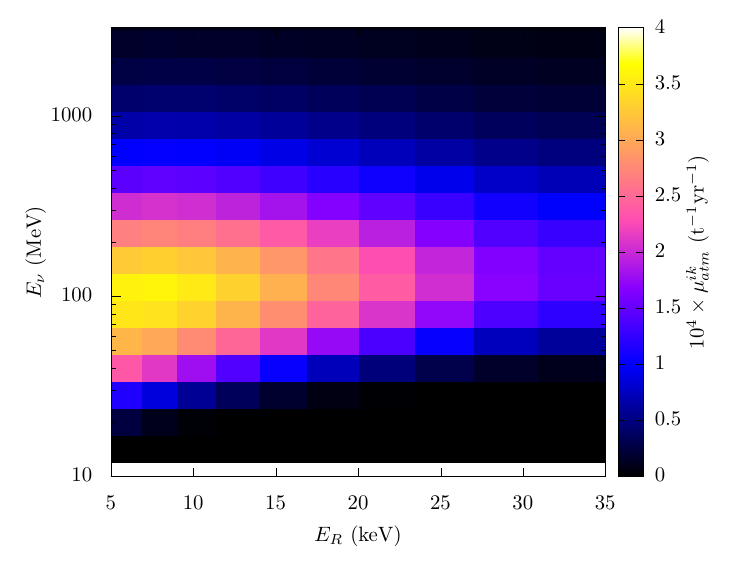}
  \includegraphics[width=.47\textwidth]{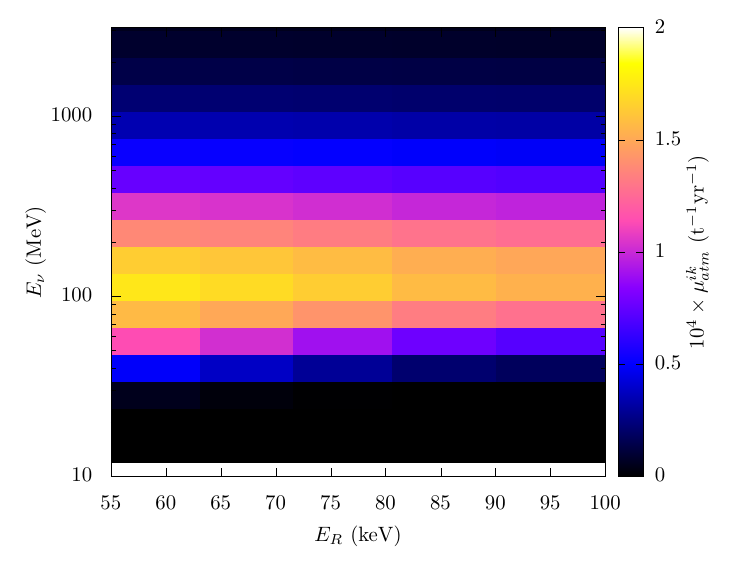}
  \caption{Atmospheric neutrino scattering event rates per detector mass $\mu^{ik}$ within the $i$-th nuclear recoil energy bin and the $k$-th atmospheric neutrino energy bin. Rates for xenon are shown in the left panel and rates for argon are shown on the right.}
  \label{fig:atmcont}
\end{figure}

 In Fig.~\ref{fig:atmcont} we show the partial counts per unit exposure $\mu^{ik}$ for xenon~(left) and argon~(right). The detector parameters used are the same as used previously and summarized in Table~\ref{tab:detector}. This figure indicates that the dominant contribution to nuclear scattering from atmospheric neutrinos comes from the neutrino energy range $E_\nu \sim 50$--$200\,\mev$. This is largely due to the flux peaking in this range, but also from the kinematics of the scattering covered in Sec.~\ref{sec:nson}.

To estimate the impact of atmospheric spectral shape uncertainties on the dark matter discovery sensitivity of future detectors using the profile likelihood framework, we need a likelihood function for variations in the shape function. Recall that previously we only varied the overall atmospheric flux normalization with a Gaussian likelihood weight given in Eq.~\eqref{eq:bglhood}. If we had a motivated functional parametrization for the atmospheric spectral, it would be straightforward to generalize this likelihood function to include variations in the model parameters. However, since we do not have such a form, we pursue a different but motivated approach.
 
Our generalized likelihood function for the atmospheric flux shape is 
\beq
\mathcal{L}_{atm} = \mathcal{N}\,e^{-S_{atm}} \ ,
\eeq
with $\mathcal{N}$ a constant normalization and $S$ is given by
\beq
S_{atm} = \frac{y_{max}}{2(\Delta\theta)^2}\,\int_{0}^{y_{max}}\!\!dy\;
\left[\frac{1}{y_{max}^2}\theta^2 + Z(\del_y\theta)^2 \right] \ ,
\label{eq:atmweight}
\eeq
where $y = \ln(E_{\nu}/10\,\mev)$, $y_{max} = \ln(10\,\gev/10\mev)$, $Z$ is a constant, 
\beq
\theta = \frac{\phi_{atm}-\bar{\phi}_{atm}}{\bar{\phi}_{atm}} \ ,
\eeq
 is the local fractional variation of the atmospheric neutrino flux spectrum relative to the fiducial value from Ref.~\cite{Battistoni:2005pd}, and $\Delta\theta = 0.2$ is the estimated fractional uncertainty on the total atmospheric flux. The first term in Eq.~\eqref{eq:atmweight} reproduces the previous likelihood when $\theta$ is constant in $y$. The second term is new and depends on the constant $Z$; it imposes a penalty on local variations in $\theta(y)$. For example, for $Z=1/4$ a linear variation in $\theta(y)$ from $\theta(0)=-1$ to $\theta(y_{max})=+1$ would contribute the same likelihood cost (from the derivative term) as the global variation $\theta(y)= 1$ everywhere (from the non-derivative term). While the form of Eq.~\eqref{eq:atmweight} is somewhat arbitrary, we argue that, for $Z\sim 1$, it allows for a reasonable estimate of the impact of atmospheric flux shape uncertainties on dark matter discovery. Ultimately, a direct comparison to atmospheric flux measurements such as in Ref.~\cite{Richard:2015aua} would be preferable provided they can be extended to lower neutrino energies.

 Our strategy to implement local variations in the atmospheric flux in estimating the dark matter discovery sensitivity is to use a discretized form of $\theta(y)$. As in Eq.~\eqref{eq:enubins} above, we split the atmospheric neutrino energies into logarithmic bins $k=1,2,\ldots\,N_\nu = 20$ ranging between $E_\nu \in [10^1,10^4]\,\mev$. The contribution to the number of atmospheric neutrino events in recoil energy bin $i$ is 
\beq
N^i_{atm} = MT\,(1+\theta_k)\,\mu^{ik} \ ,
\eeq
 where $\theta_k$ is a (weighted) discretized form of $\theta_y$ that characterizes the variation in the atmospheric flux over $E_\nu \in [E_{k-1},E_k]$. These $\theta_k$ variables are then treated as nuisance parameters in the profile likelihood with a likelihood weight motivated by Eq.~\eqref{eq:atmweight} of
\beq
S_{atm} = \frac{1}{2(\Delta\theta)^2}\sum_{k=1}^{N_{\nu}}
\left[
\frac{1}{N_{\nu}}\,\theta_k^2
+ Z\,N_{\nu}\,(\theta_{k+1}-\theta_k)^2(1-\delta_{k,N_\nu})
\right] \ ,
\eeq
 where we set $Z=1/4$. To achieve a more robust minimization in the profile likelihood, we find it convenient to first minimize for rigid $\theta = \theta_1 = \theta_2 = \ldots = \theta_{N_\nu}$, and then reminimize in variations around this constrained minimum.

\begin{figure}[ttt]
  \centering
  \includegraphics[width=.47\textwidth]{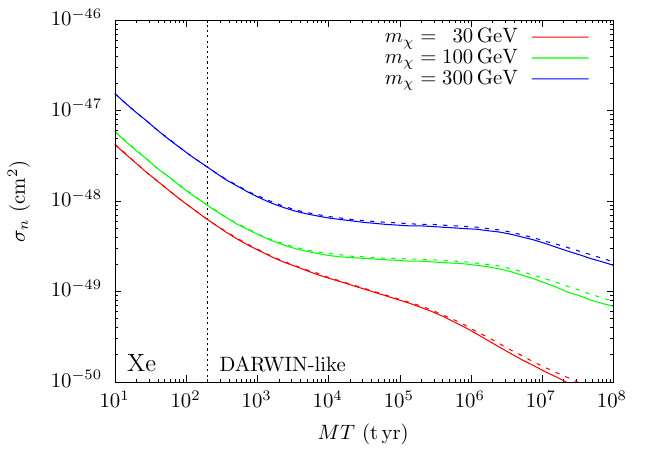}
  \includegraphics[width=.47\textwidth]{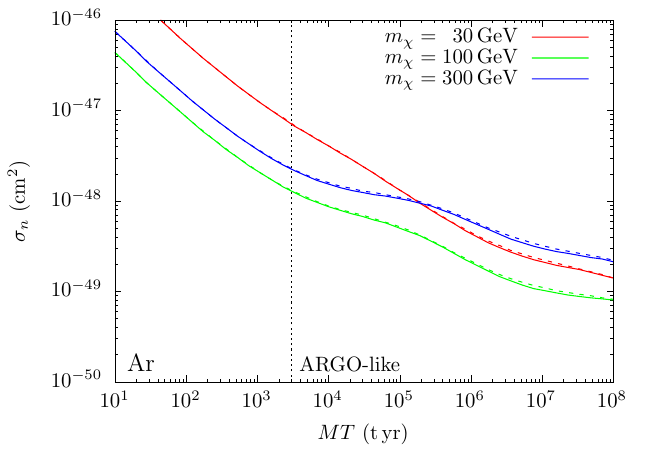}
  \caption{Atmospheric neutrino scattering event rates per detector mass $\mu^{ik}$ within the $i$-th nuclear recoil energy bin and the $k$-th atmospheric neutrino energy bin. Rates for xenon are shown in the left panel and rates for argon are shown on the right.}
  \label{fig:mt-fratm}
\end{figure}

 The results of this approach for the dark matter discovery limits in xenon and argon are shown in Fig.~\ref{fig:mt-fratm}. Allowing for an increased freedom in the spectral shape of the atmospheric neutrino flux degrades the sensitivity to dark matter, but only by a very small amount. This indicates that allowing the normalization of this flux source to float captures most of the uncertainty on it. The result is also consistent with the doubly differential event rates shown in Fig.~\ref{fig:atmcont}, where we see that the partial atmospheric event counts in each nuclear recoil energy bin are dominated by atmospheric neutrinos with energies close to $E_{\nu} \sim 100\,\mev$.

\bibliography{BibFile}

\end{document}